\documentclass[sigconf,screen]{acmart}
\usepackage{rotating}
\usepackage{adjustbox}
\usepackage{xspace}
\usepackage{multirow}
\usepackage{enumitem}
\usepackage{subcaption}
\usepackage{hyperref}
\usepackage{graphicx}
\usepackage{listings}
\usepackage[many]{tcolorbox} 
\usepackage{booktabs}
\usepackage{rotating}
\usepackage{pdflscape}
\usepackage{xurl}
\usepackage{seqsplit}
\usepackage[htt]{hyphenat}
\usepackage{microtype}
\usepackage{url}

\newtcolorbox{boxE}{
    enhanced, 
    boxrule = 0pt, 
    borderline = {0.75pt}{0pt}{main}, 
    borderline = {0.75pt}{2pt}{sub} 
}
\newtcolorbox{boxF}{
    colback = sub,
    enhanced,
    boxrule = 1.5pt, 
    colframe = white, 
    borderline = {1.5pt}{0pt}{main, dashed} 
}
\usepackage{circledsteps}

\usepackage{amsmath}
\usepackage{pifont}

\usepackage{enumitem}
\setlist{noitemsep, topsep=2pt, parsep=0pt, partopsep=0pt}

\definecolor{codegreen}{rgb}{0,0.6,0}
\definecolor{codegray}{rgb}{0.5,0.5,0.5}
\definecolor{codepurple}{rgb}{0.58,0,0.82}
\definecolor{backcolour}{rgb}{0.95,0.95,0.92}
\definecolor{codeblue}{rgb}{0.0,0.0,0.8}

\newcommand{\hyph}{\nobreak-\hspace{0pt}\allowbreak}

\newcommand{\mypara}[1]{\vspace{0.3em}\noindent\textbf{#1}\hspace{0.1em}}

\newcommand{\ie}{\emph{i.e.,}\ } 

\newcommand{\chatgptfouro}{\texttt{GPT\hyph 4o}\xspace}
\newcommand{\geminionefiveflash}{\texttt{Gemini\hyph 1.5\hyph Flash}\xspace}
\newcommand{\geminionefivepro}{\texttt{Gemini\hyph 1.5\hyph Pro}\xspace}
\newcommand{\geminitwofiveflash}{\texttt{Gemini\hyph 2.5\hyph Flash}\xspace}
\newcommand{\chatgptthreefive}{\texttt{GPT\hyph 3.5\hyph Turbo}\xspace}
\newcommand{\chatgptfouromini}{\texttt{GPT\hyph 4o\hyph mini}\xspace}
\newcommand{\claudesonnet}{\texttt{Claude\hyph 3\hyph Sonnet}\xspace}
\newcommand{\claudesonnetthreefive}{\texttt{Claude\hyph 3.5\hyph Sonnet}\xspace}
\newcommand{\gemini}{\texttt{Gemini\hyph 1.5\hyph Flash}\xspace}

\newcommand{\fewshot}{\texttt{Few\hyph Shot}\xspace}

\newcommand{\minpro}{\textsc{Minimal}\xspace}
\newcommand{\detpro}{\textsc{Detailed}\xspace}
\newcommand{\pypro}{\textsc{Python}\xspace}

\newcommand{\eg}{\emph{e.g.}\xspace}

\newcommand{\llmft}{\ensuremath{\mathtt{LLM_{ft}^{RI/O}}}\xspace}
\newcommand{\llmgp}{\ensuremath{\mathtt{LLM_{gp}}}\xspace}

\newcommand{\nlltltask}{\ensuremath{\mathtt{NL}\!\!\!\rightarrow\!\!\mathtt{LTL}}\xspace}

\newcommand{\nl}{\ensuremath{\mathtt{NL}}\xspace}
\newcommand{\ltl}{\ensuremath{\mathtt{LTL}}\xspace}
\newcommand{\pl}{\ensuremath{\mathtt{PL}}\xspace}

\newcommand{{\nusmv}}{\ensuremath{\mathtt{NuSMV}}\xspace}

\newcommand{\gt}{\ensuremath{\mathtt{GT}}\xspace}
\newcommand{\pred}{\ensuremath{\mathtt{Pred}}\xspace}

\newcommand{\addformula}[1]{\texttt{\seqsplit{#1}}\xspace}

\newcounter{researchquestion}[section]

\newcommand{\rqtype}[2]{\ensuremath{\mathbb{RQ}^\mathsf{#1}_\mathbf{#2}}\xspace}
\newcommand{\RQtype}[3]{\rqtype{#1}{#2}: \emph{#3}}

\newcommand{\dstype}[2]{\ensuremath{\mathbb{DS}^\mathsf{#1}_\mathbf{#2}}\xspace}

\newcommand{\extype}[3]{\ensuremath{\mathbb{E}_\mathtt{#1}{#2} \rightarrow \textbf{#3}}\xspace}
\newcommand{\Exptype}[3]{\extype{#1}{#2}{#3}:}

\setcopyright{cc}
\setcctype{by-nc-nd}
\acmDOI{10.1145/3805773.3806005}
\acmYear{2026}
\copyrightyear{2026}
\acmISBN{979-8-4007-2602-6/2026/07}
\acmConference[SecDev '26]{Proceedings of the 2026 ACM Secure Development Conference}{July 5--6, 2026}{Montreal, QC, Canada}
\acmBooktitle{Proceedings of the 2026 ACM Secure Development Conference (SecDev '26), July 5--6, 2026, Montreal, QC, Canada}
\acmSubmissionID{fsesecdev26main-p38-p}
\received{2026-02-05}
\received[accepted]{2026-03-19}
\begin{document}

\title{Syntax Is Easy, Semantics Is Hard: \\ Evaluating LLMs for LTL Translation}

\author{Priscilla Kyei Danso}
\orcid{0000-0002-1262-844X}
\affiliation{%
  \institution{Stony Brook University}
  \city{Stony Brook}
  \country{USA}
}
\email{priscilla.danso@stonybrook.edu}

\author{Mohammad Saqib Hasan}
\orcid{0000-0001-7588-8591}
\affiliation{%
  \institution{Stony Brook University}
  \city{Stony Brook}
  \country{USA}
}
\email{mdshasan@cs.stonybrook.edu}

\author{Niranjan Balasubramanian}
\orcid{0000-0003-4187-9368}
\affiliation{%
  \institution{Stony Brook University}
  \city{Stony Brook}
  \country{USA}
}
\email{niranjan@cs.stonybrook.edu}

\author{Omar Chowdhury}
\orcid{0000-0002-1356-6279}
\affiliation{%
  \institution{Stony Brook University}
  \city{Stony Brook}
  \country{USA}
}
\email{omar@cs.stonybrook.edu}

\renewcommand{\shortauthors}{Danso et al.}

\begin{abstract}
Propositional Linear Temporal Logic (LTL) is a popular formalism 
for specifying desirable requirements and security and privacy 
policies for software, networks, and systems. 
Yet expressing such requirements and policies in LTL remains 
challenging because of its intricate semantics. 
Since many security and privacy analysis tools require LTL formulas 
as input, this difficulty places them out of reach for many developers 
and analysts. 
Large Language Models (LLMs) could broaden access to such tools by 
translating natural language fragments into LTL formulas. 
\emph{This paper evaluates that premise by assessing how effectively 
several representative LLMs translate assertive English sentences into 
LTL formulas.} 
Using both human-generated and synthetic ground-truth data, we evaluate 
effectiveness along syntactic and semantic dimensions. 
The results reveal three findings: (1) in line with prior findings, 
LLMs perform better on syntactic aspects of LTL than on semantic ones; 
(2) they generally benefit from more detailed prompts; and 
(3) reformulating the task as a Python code-completion problem 
substantially improves overall performance. 
We also discuss challenges in conducting a fair evaluation on this task and conclude with recommendations for future work. 
\end{abstract}


\maketitle

\section{Introduction} \label{sec:intro}

Combining Large Language Models (LLMs) and symbolic analysis \& reasoning (\textbf{\textit{SR}})~\cite{yao2024chainofthoughteffectivegraphofthoughtreasoning, Kambhampati_2024} is often mutually beneficial: (i) SR tools can enhance LLMs’ symbolic reasoning capabilities and reduce hallucinations \cite{zhou2023leasttomostpromptingenablescomplex}; and (ii) LLMs can substantially improve the usability of SR tools by providing their users with a natural language (\nl) interface \cite{besold2017neuralsymboliclearningreasoningsurvey}. Enabling this synergy often requires LLMs to translate assertive \nl sentences into formal logic. \emph{In this paper, we evaluate the effectiveness of representative existing LLMs in translating assertive \nl sentences into one such formalism, namely propositional Linear Temporal Logic (LTL)}, a task we call \nlltltask.

In the formal methods,  requirements engineering, security and privacy communities, 
\ltl has gotten a lot of traction as a  
natural logic of choice for formally expressing guarantees, security and privacy 
policies, security and privacy properties, and 
requirements of systems, software, and networks that are amenable to automatic analysis~\cite{Pnueli_1977}, 
runtime monitoring \cite{10.1145/2000799.2000800}, testing \cite{10.1145/1291535.1291542}, and synthesis \cite{F16}. 
Different fragments of LTL have been also used in many SR tools  relevant to the privacy and security communities: 
access control policies and their analyses \cite{accesscontrol1,accesscontrol2,maverick,cloudfix}, 
program repair \cite{repair1,autotap,repair2}, 
privacy policies \cite{precis,privacypromise,eunomia}, 
safety policies and their analyses \cite{arv,crv}, 
attack detection \cite{syslite,phoenix}, 
threat modeling \cite{threatmodeling2024}, 
and 
property-guided testing and analysis \cite{quickstrom,ltlgreybox,lteinspector,5greasoner,chiron,proteus,dikeue,uptane,esim}.

Despite its popularity, manually expressing 
\nl requirements in \ltl can be challenging due to inherent ambiguity 
in natural languages \cite{Greenman_2022} and intricacies in \ltl semantics. This task is shown 
to be challenging even for experts, let alone regular users of SR tools 
\cite{Greenman_2022}. Hence, one of the main impediments towards wide adoption of 
many SR tools (\eg, for security/privacy analysis) that use \ltl as a specification 
language \cite{10.1007/s00165-019-00486-z} is the manual step of \nlltltask. 

To reduce users’ burden, prior work has tried to ease this error-prone manual step \cite{10.1145/302405.302672, 1553580}. One line of work studies properties from the literature and identifies recurring patterns to help users systematically, though still manually, translate \nl requirements into \ltl \cite{10.1145/302405.302672}. Other efforts develop automated tools for translating restricted fragments of \nl requirements into \ltl \cite{10.1007/11901433_41, Farrell_2024}. Despite this progress, interest remains in translating free-form \nl requirements into \ltl and other logics \cite{10.1007/978-3-540-30206-3_12, 10.5555/332656}.

LLMs hold promise for automating the error-prone and challenging process of writing \ltl specifications through \nlltltask. Beyond proprietary general-purpose LLMs \cite{rahman2025comparativeanalysisbaseddeepseek, RAY2023121}, some efforts have fine-tuned open-source models for this task \cite{pirozelli2024assessinglogicalreasoningcapabilities}. \emph{Although these approaches have shown promise, the literature still lacks a systematic evaluation of their effectiveness on \nlltltask and a definitive answer to whether LLMs can enable broader use of these analysis tools by developers and general users.}

This paper presents such an in-depth evaluation. We first collect \nlltltask ground-truth datasets from multiple sources, including textbooks \cite{NEURIPS2023_91edff07,fisher2011introduction,jeffrey2006formal}, papers \cite{Greenman_2022}, and manually constructed examples \cite{8972130}, and then evaluate representative general-purpose LLMs from two perspectives: the \emph{syntax} and \emph{semantics} of \ltl. We assemble these datasets to ensure that the test set captures organically occurring requirements rather than contrived ones. We note that some LLMs perform \nlltltask through a collaborative human-in-the-loop process \cite{brown2020languagemodelsfewshotlearners}, whereas others perform the translation fully automatically, without substantial human intervention \cite{10684640}. \emph{This paper primarily focuses on evaluating LLMs in the latter category.}

Conceptually, to evaluate an LLM’s efficacy on a ground-truth entry $\langle s: \nl, f_g : \ltl \rangle$, one would feed the \nl sentence $s$ (with appropriate prompting text) to the LLM under test and obtain an \ltl formula $f$. One could then check whether the output matches the ground truth by testing logical equivalence, that is, $f \equiv^? f_g$. Unfortunately, this natural high-level approach faces a major challenge: the \emph{ontological problem of choosing the same set of propositions} as the ground truth. Concretely, checking $f \equiv^? f_g$ is not meaningful when $f$ and $f_g$ use different sets of propositional variables.

This problem is further exacerbated for fine-tuned LLMs with a restricted input-output interface (\llmft), which take an \nl sentence and output an  \ltl formula. This precludes using prompting text to provide a fixed mapping from \nl fragments to propositional variables. Thus, these \llmft{}s are not amenable to fully automated, large-scale evaluation. A natural follow-up question is whether such \llmft{}s do map the same \nl fragments to the corresponding ground-truth propositional variables, but merely use different variable names. 
If so, it is conceivable to devise an automatic \emph{variable renaming} scheme to address the ontological problem.

Consequently, we constructed a ground-truth \nl-to-propositional-logic (\pl) dataset (recall that \pl is a sublogic of \ltl) and manually evaluated the efficacy of the \llmft{}s under test in selecting the correct \nl fragments to map to propositional variables. Unfortunately, we found that these \llmft{}s violate the \emph{principle of maximal logical revelation} (\ie always translate so as to reveal as much logical structure as the target language supports) \cite{Greenman_2022}. This issue is particularly relevant when the model of the system, software, or network being analyzed or monitored is at a different level of abstraction (\eg, more fine-grained) than the property being analyzed or monitored on that model (\eg, more coarse-grained), since such a mismatch between the vocabularies of the model and the property can make the analysis inapplicable. We observed similarly unsatisfactory performance from proprietary general-purpose LLMs, which do not have such a restricted input-output interface and which we refer to as \llmgp, especially when they are not explicitly instructed in the prompt to respect this principle for proposition selection. Fortunately, prompting can provide a fixed mapping to \llmgp{}s and thereby sidestep this ontological problem. Although many models still do not strictly adhere to the provided mapping when choosing propositions, approaches such as grammar-constrained decoding can address this problem \cite{gcd}.

In terms of features, we observe that almost all \llmft{}s support only the future-only fragment of \ltl. Although adding past temporal operators (\ie since and yesterday) to \ltl does not increase its expressive power, it allows certain formulas to be expressed more succinctly \cite{pastltl}. Moreover, the past-only fragment of \ltl captures monitorable safety properties under the standard finite-trace semantics and is therefore a popular choice for specifying security properties \cite{geatti_et_al:LIPIcs.TIME.2021.8,patriot,maverick}. We also observe that almost all \llmgp{}s perform better on the future-only fragment than on the past-only fragment of \ltl. Finally, because of their restricted input-output interface, \llmft{}s do not support many of the syntactic aspects of our evaluation, nor some semantic aspects. Overall, for \llmgp{}s, we find that they perform substantially better on syntactic aspects than on semantic ones, and that they improve markedly when given a detailed prompt, possibly with examples, that covers the nuanced aspects of the task, consistent with conventional wisdom.

The most notable finding of our evaluation is that 
reformulating the \nlltltask task to a Python code 
completion task substantially improves \llmgp{}'s overall 
performance. Despite this substantial 
improvement in performance, one surprising fact was that 
in one of the experiments about LTL semantics, which essentially boils down to  
executing a Python program, the \llmgp{}s, despite being equipped with 
Python interpreters, did not achieve full accuracy. 
To further assess security relevance, we also conduct an end-to-end evaluation on 56 randomly selected security-focused requirements from the VERIFY  dataset~\cite{quansah2026verify}, without providing a fixed atomic proposition (AP) mapping. These requirements capture security-relevant behaviors such as authentication persistence, session validity, re-authentication triggers, and policy-driven access control. The results show that, although the security domain is modestly harder and correct AP grounding becomes more important, temporal operator mis-scoping remains the dominant source of error.

We design and implement a novel six-tier evaluation framework for \nlltltask translation, introduce three interfaces (\ie prompts) with progressively stronger structural constraints, and employ NuSMV as an automated semantic oracle enabling principled, reproducible assessment at both syntactic and semantic levels.
To summarize, the paper has the following contributions: 
\begin{enumerate}
    \item Using this framework, we develop the first systematic and in-depth evaluation of LLMs' efficacy on the \nlltltask translation from both syntactic and semantic aspects of \ltl. 
    \item We identify rooms for improvement of \llmft for the \nlltltask task. 
    \item We observe that \llmgp{}s substantially perform better 
    when the \nlltltask task is reformulated as Python code completion and comprehension task. 
    
    \item We also present technical
challenges of having a fair evaluation and conclude with
recommendations for future evaluations of LLMs, especially developed for the \nlltltask task.
\end{enumerate}

\noindent\textbf{Artifacts.}
All artifacts are available in the  Github repository \cite{prompteval}.

\section{Motivation}
\label{sec:motivation}
This section outlines the research questions driving our study and their motivations. 
Our evaluation goes beyond the standard end-to-end \nlltltask task, and for \llmgp{}s also aims to gauge their proficiency 
in answering questions about syntactic (\eg, well-formedness) and 
semantic (\eg, trace classification) aspects of \ltl. 

\mypara{Motivation for our experimental setup.} 
Consider a fictitious setting in which a researcher is interested 
in using a property-guided testing approach on a protocol implementation to discover vulnerabilities in it \cite{quickstrom,ltlgreybox,lteinspector,5greasoner,chiron,proteus,dikeue,uptane,esim}. 
However, one of the main challenges of applying one of these testing approaches is that 
they expect the properties that guide the testing, whose violation entails a logical vulnerability in the implementation, 
to be expressed in \ltl. Unfortunately, the researcher finds the semantics 
of \ltl to be unintuitive. To make matters worse, the researcher only has the 
protocol specification in the form of a natural language document (\ie an RFC document). 
After consulting the RFC, the researcher realizes that the RFC does not explicitly 
enumerate all the properties the protocol design guarantees in a well-marked section. 
To apply the above property-guided testing, the researcher has to achieve the following. 
\ding{182} The researcher first has 
to identify the \nl fragments from the RFC that correspond to a desirable 
guarantee, whose violation can be exploitable by the adversary. 
\ding{183} For each of the identified \nl fragments that corresponds to a desirable guarantee, they 
then have to convert those \nl fragments to the corresponding \ltl formula. In our evaluation 
and experiments, we focus mainly on step \ding{183}. Although one may 
argue that our experimental setup is not organic, we argue that if the LLMs 
cannot even perform step \ding{183} correctly, then they are unlikely to be 
successful is performing the end-to-end task consisting of both steps \ding{182} and \ding{183}. 
\emph{In that sense, our evaluation can be viewed as a lower-bound of the original end-to-end 
task of identifying \nl fragments that correspond to properties, and translate them to their corresponding 
\ltl formulas.}

\mypara{Motivation for Evaluating LLMs' \ltl Reasoning Capabilities.}
A natural question readers may have is that when there are 
symbolic reasoners available for carrying out different reasoning 
tasks (\eg, model checkers, reactive synthesizer) what is the point 
of using LLMs for such semantic reasoning tasks (See Section \ref{subsec:sem_eval_question}). There have been a 
growing interests in using LLMs for different temporal reasoning tasks 
(\eg, satisfiability \cite{LLMLTLSAT}, model checking \cite{LLMMC1,LLMMC2}, reactive synthesis \cite{LLMSYNTH}). Our evaluation 
aims to answer whether the LLMs have a basic understanding of the \ltl semantics.
In case LLMs cannot correctly answer simple \ltl semantics-related queries (\eg, trace 
characterization), then it is unlikely to faithfully perform more complex semantic 
reasoning, which have substantial theoretical complexity (\eg, PSPACE-complete for 
\ltl satisfiability).

\subsection{Research Questions}
This section presents the research questions guiding our evaluation, categorized 
into both syntactic and semantic categories. 

\mypara{Syntactic Evaluation:} We have the following syntactic category of research questions. 

\RQtype{syn}{1}{How effectively can LLMs extract atomic propositions from natural language requirements?}  
This question investigates whether LLMs can identify the key propositional elements in a natural language requirement. Success here reflects the model's ability to capture the minimal syntactic building blocks necessary for further logical reasoning.

\RQtype{syn}{2}{How accurately can LLMs distinguish well-formed from malformed \ltl formulas?}  
This question examines whether LLMs understand the syntactic structure of \ltl formulas, including the correct use of connectives, modalities, and parentheses. 
This is analogous to parsing an \ltl formula. 

\mypara{Semantic Evaluation:} Our evaluation of the semantic aspects of \ltl is driven 
by the following research questions. 
\label{subsec:sem_eval_question}

\RQtype{sem}{1}{Can LLMs generate LTL formulas that are semantically equivalent to the intended ground truth?}  
This question focuses on whether LLMs can produce formulas that, despite possible syntactic differences, preserve the same meaning as a reference specification.

\RQtype{sem}{2}{Can LLMs accurately assess whether a trace satisfies a given formula?}  
This investigates the LLMs' ability to reason about traces, determining whether sequences of system behaviors satisfy or violate a given specification.

\RQtype{sem}{3}{Can LLMs generate satisfiable traces for a given \ltl formula?}  
This question explores whether LLMs can generate traces consistent with the constraints of a formula, reflecting an understanding of global temporal dependencies.

\RQtype{sem}{4}{Can LLMs generate semantically equivalent \textbf{Past}-\ltl formulas from natural language descriptions?}  
This question addresses whether LLMs can correctly translate natural language involving past temporal operators into formulas equivalent to past-LTL.

\section{The Evaluation Framework}
\label{sec:framework}

We describe the evaluation framework employed throughout this work. It assesses the semantic correctness of LLM-generated temporal specifications with respect to formal trace semantics, decomposing the \nlltltask translation pipeline into well-defined evaluation tasks regulated through carefully designed interfaces.

\subsection{Experiments \label{sub:experiments}}
This section describes all the experiments we carry out. 
Table~\ref{tab:research-mapping} summarizes the relationship between research questions, datasets, metrics, and oracles across tasks targeting syntactic competence, semantic equivalence, and trace reasoning.

\Exptype{nl2pl}{(NL)}{List(NL phrase$\mapsto$ AP)} Given NL text, models extract atomic propositions mapped to formal variables. Evaluation uses Levenshtein distance and Jaccard similarity.

\Exptype{wff}{(formula)}{\{Yes, No\}} Given an LTL formula, models classify it as well-formed or ill-formed. Evaluation uses accuracy, precision, recall, and F1-score.

\Exptype{nl2ltl}{(NL, AP)}{LTL Formula} Given NL and fixed AP mappings, models generate future-time LTL formulas across all three prompting strategies. For \textsc{Python}, models output Python AST representations. Semantic correctness is evaluated via equivalence and one-way entailment checks using a symbolic verification backend.

\Exptype{tracechar}{(LTL, Trace)}{\{Yes, No\}} Given an LTL formula and execution trace, models classify whether the trace satisfies the formula. Predictions are evaluated using accuracy and F1-score.

\Exptype{tracegen}{(formula)}{($\tau^+/\tau^-$)} Models generate satisfying and violating traces for a given LTL formula, automatically verified using a formal checker.

\Exptype{nl2pltl}{(NL, AP)}{Past LTL Formula} Mirroring future LTL generation, models generate past-time LTL formulas. Semantic correctness is evaluated via equivalence and entailment checks against ground truth.

\begin{table*}[!htbp]
\caption{Task–evaluation mapping across experiments, research questions, datasets, and interfaces.}
\label{tab:research-mapping}
\centering
\scalebox{0.8}{
\begin{tabular}{|c|c|c|c|c|c|c|c|c|}
\hline
\textbf{RQ} & \textbf{Task} & \textbf{Input} & \textbf{Output} & \textbf{Metric} & \textbf{Dataset} & \textbf{Size} & \textbf{Interfaces} & \textbf{Oracle} \\ \hline
RQ$_{\text{syn}}^{1}$ & Atomic proposition extraction & NL & AP set & Levenshtein, Jaccard & \dstype{prop}{5} & 144 & Minimal, Detailed & Manual \\ \hline
RQ$_{\text{syn}}^{2}$ & Well-formedness & LTL & Yes/No & Accuracy, F1 & \dstype{syntax}{6} & 299 & Minimal, Detailed & Grammar \\ \hline
RQ$_{\text{sem}}^{1}$ & NL$\rightarrow$Future LTL & NL + AP & LTL & Equiv., Entail. & \dstype{tricky}{1}, \dstype{book}{3} & 306, 141 & Minimal, Detailed, Python & NuSMV \\ \hline
RQ$_{\text{sem}}^{2}$ & Trace classification & LTL + trace & Yes/No & Accuracy, F1 & \dstype{trace}{4} & 306 & Minimal, Detailed, Python & NuSMV \\ \hline
RQ$_{\text{sem}}^{3}$ & Trace generation & LTL & $(\tau^+/\tau^-)$ & Trace satisfaction & \dstype{trace}{4} & 306 & Minimal, Detailed, Python & NuSMV \\ \hline
RQ$_{\text{sem}}^{4}$ & NL$\rightarrow$Past LTL & NL + AP & Past-LTL & Equiv., Entail. & \dstype{pastltl}{2} & 294 & Minimal, Detailed, Python & NuSMV \\ \hline
\end{tabular}
}
\end{table*}

\subsection{Interface Design}
\label{subsec:interfaces}
We evaluate three interfaces with increasing constraint: minimal prompts, detailed prompts including grammar and semantics, and a Python-based AST interface enforcing compositional structure. 
Detailed prompt structure for each interface is in Appendix~\ref{appendix:prompts}.

\noindent\textbf{Minimal Interface.} Models receive brief task descriptions and minimal symbolic definitions with no formal grammars, relying on implicit knowledge of LTL syntax and semantics.

\noindent\textbf{Detailed Interface.} Prompts include full BNF grammar, semantic definitions, operator constraints, and examples, assessing whether formal context alone supports reliable semantic reasoning.

\noindent\textbf{Python (AST) Interface.}
Formulas are constructed as Python dataclass abstract syntax trees (ASTs) and traces as Boolean sequences.
This representation enforces syntactic validity and compositional structure during generation by mapping each logical and temporal operator to a dedicated constructor with fixed semantics.
The AST representation constrains operator composition and reduces ill-formed outputs.
AST-generated formulas are deterministically converted into equivalent LTL strings prior to any semantic equivalence or entailment checks to use with \nusmv{} which operates with symbolic LTL syntax.
Thus, the AST functions as a structured intermediate representation that guides generation while preserving compatibility with established symbolic verification workflows.
Performance gains may stem from both LLM familiarity with Python code and the structural scaffolding that reduces operator misuse.



\subsection{Role of NuSMV as a Semantic Oracle}

We use NuSMV as an external semantic oracle for equivalence checking, entailment, and trace validation.
Model-generated formulas and traces are evaluated through model-checking queries, ensuring objective semantic judgments grounded in established verification practice.
Together, constrained interfaces and oracle-backed evaluation enable principled and reproducible assessment of LLMs as semantic interfaces for temporal specification.

\section{Experimental Setup}
\label{sec:experimental_setup}

This section instantiates the evaluation framework of Section~\ref{sec:framework}, describing datasets, prompting strategies, models, and evaluation metrics. Evaluating data contamination is beyond this work's scope, but we provide dataset details for reproducibility.

\subsection{Datasets}
\label{subsec:datasets}

We curated multiple ground-truth datasets from diverse sources to enable comprehensive evaluation.

\noindent\textbf{\dstype{tricky}{1} (Future-Time LTL).} Derived from the \emph{Little Tricky Logic} benchmark~\cite{Greenman_2022}, this dataset contains 306 instances, each comprising a natural-language requirement, a ground-truth future-time LTL formula, and an AP mapping. It captures recurring sources of semantic error including nested temporal scopes, conditional obligations, and implicit precedence, and serves as the primary benchmark for future-time \nlltltask translation.

\noindent\textbf{\dstype{pastltl}{2} (Past-Time LTL).} Adapted by rephrasing requirements using past-time constructions, this dataset contains 294 instances paired with ground-truth past-time LTL formulas and AP mappings. It isolates the effect of temporal directionality on semantic correctness and enables comparison with future-time reasoning.

\noindent\textbf{\dstype{book}{3} (Textbook LTL).} Collated from logic and formal specification textbooks~\cite{NEURIPS2023_91edff07,fisher2011introduction,jeffrey2006formal,8972130}, this dataset contains 141 NL-formula pairs with derived AP mappings, covering complex real-world specifications with formula complexity dependent on the NL description.

\noindent\textbf{\dstype{trace}{4} (Trace Reasoning).} For each formula in \dstype{tricky}{1}, satisfying and violating finite traces are generated using NuSMV. Traces are randomized to avoid positional bias and serve as ground truth for trace characterization and generation tasks, enabling semantic evaluation independently of formula generation.

\noindent\textbf{\dstype{prop}{5} (Propositional Logic).} Containing 144 examples from textbooks and formal reasoning exercises, each entry includes an NL statement, a ground-truth propositional formula, and an AP mapping. Used solely for atomic proposition extraction, isolating phrase extraction from temporal reasoning.

\noindent\textbf{\dstype{syntax}{6} (Well-Formedness).} A set of 299 LTL formulas sampled from a context-free grammar, containing both well-formed and ill-formed instances with AST depths ranging from atomic propositions to deeply nested constructs. Used exclusively for well-formedness classification.

\subsection{Prompting Techniques}
\label{prompts}

We evaluate three prompting techniques orthogonal to the interface designs of Section~\ref{subsec:interfaces}.

\noindent\textbf{Zero-Shot Prompting.} Models receive only a minimal task description with no examples, evaluating intrinsic LTL reasoning from pretraining alone. Models are instructed to generate answers without explanation.

\noindent\textbf{Few-Shot Prompting.} Models are given 3 fixed annotated examples illustrating the task format, evaluating whether limited structural cues improve performance.

\noindent\textbf{Zero-Shot Self-Refinement.} Extending zero-shot prompting with a single self-guided revision step, the model refines its initial output using only its prior response—no external feedback or oracle signal is provided. This assesses whether models can internally identify and correct semantic errors from parametric knowledge alone.

\subsection{Models}
\label{subsec:models}

We evaluate \path{GPT-3.5-Turbo}, \path{GPT-4o}, \path{GPT-4o-Mini}, \path{Gemini-1.5-Flash}, \path{Gemini-1.5-Pro}, \path{Gemini-2.5-Flash}, and \path{Claude-3.5-Sonnet}. These models differ in architecture, scale, and training, enabling comparison across diverse design choices; cost-efficient models were included to balance performance with experimental scalability. All are off-the-shelf general-purpose LLMs (\llmgp); no fine-tuning is performed, and the \llmft{} distinction serves only to contextualize challenges in constructing large-scale corpora.

\subsection{Evaluation Metrics}
\label{subsec:metrics}

\noindent\textbf{Atomic Proposition Extraction.} Evaluated using Jaccard similarity (lexical overlap) and Levenshtein distance (surface-form variation preserving semantic intent), with final correctness verified through manual inspection.

\noindent\textbf{Classification Tasks.} Accuracy, precision, recall, and F1-score are reported for well-formedness and trace characterization, with F1 emphasized under class imbalance.

\noindent\textbf{Semantic Equivalence and Entailment.} For NL$\rightarrow$LTL translation, correctness is evaluated via bidirectional entailment between the predicted formula $\varphi_{\text{pred}}$ and ground-truth $\varphi_{\text{gt}}$:
$$\varphi_{\text{pred}} \Rightarrow \varphi_{\text{gt}} \quad \text{and} \quad \varphi_{\text{gt}} \Rightarrow \varphi_{\text{pred}}$$
All checks are discharged through the NuSMV model checker.

\noindent\textbf{Trace Generation.} Predicted traces are validated against the target LTL formula using NuSMV.

\subsection{Verification Procedure}
\label{subsec:verification}

Semantic correctness for NL$\rightarrow$LTL tasks is evaluated by comparing LLM-generated formulas against ground truth via bidirectional entailment queries operating strictly at the formula level. Trace-based tasks are evaluated separately using the same oracle. Verification queries that fail due to malformed formulas or AP mismatches are recorded as Not Meaningful (N/M), excluded from accuracy calculations but reported separately.


\section{Experimental Evaluation}
\label{sec:experiments}

In this section, we present the experimental results from evaluating the performance of various LLMs' prediction (\pred) against the ground-truth (\gt) by running the experiments presented in Section~\ref{sub:experiments} using the three prompting interfaces discussed in Section~\ref{prompts}. 

\noindent
\textbf{\RQtype{syn}{1}{How effectively can LLMs extract atomic propositions from natural language?}}

\begin{figure}[!thbp]
\vspace{2mm}

\centerline{\includegraphics[width=0.99\columnwidth]{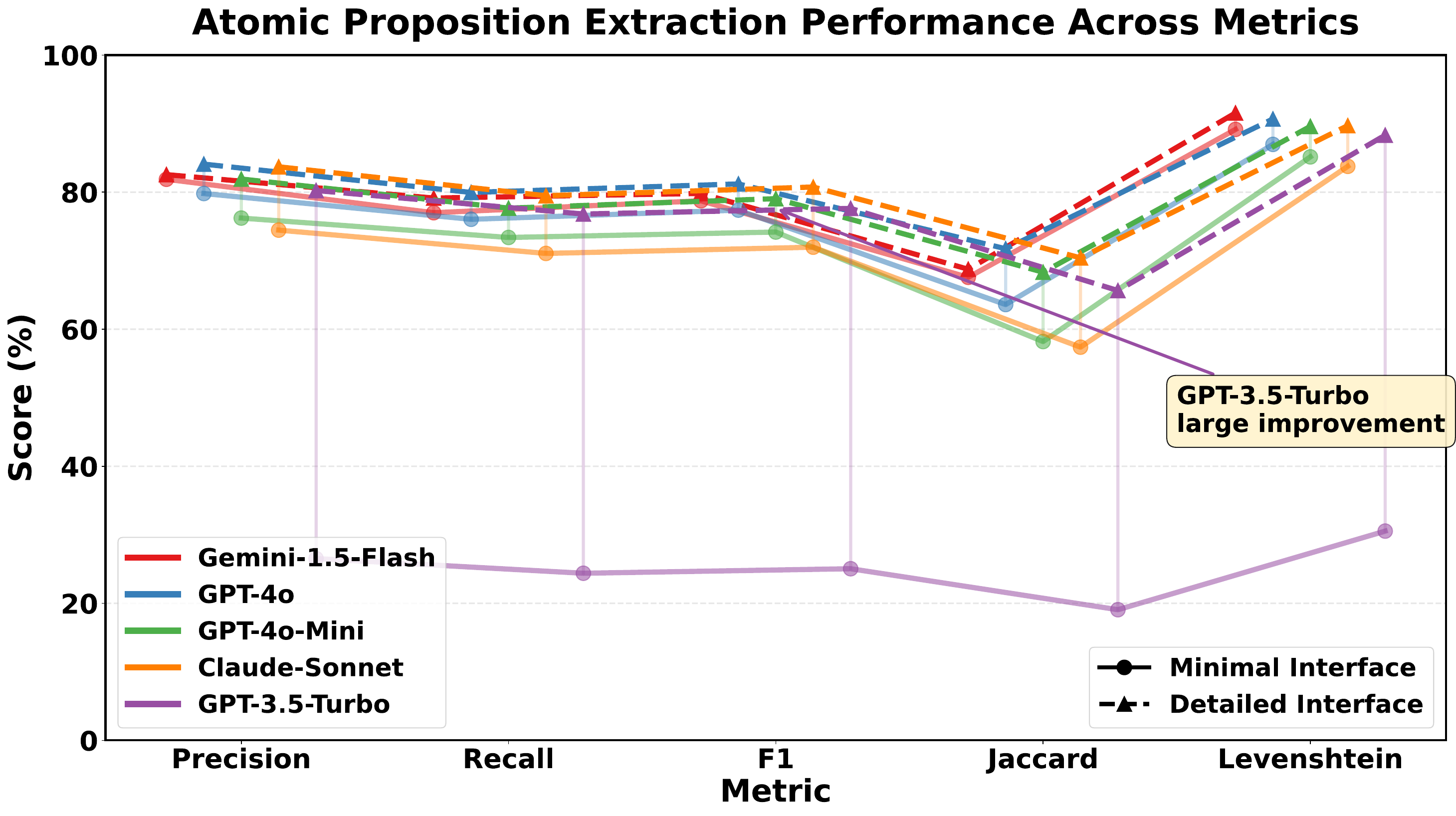}}

\Description{Line chart comparing Detailed vs. Minimal interface performance across F1, precision, recall, Jaccardi, and Levenshtein metrics.}
\caption{Performance per model and interface (precision, recall, F1-score, Jaccard similarity, Levenshtein similarity) for Atomic Proposition Extraction.}
\label{fig:ap_extraction}
\end{figure}

\noindent
\textbf{Answer.}
LLMs extract atomic propositions \textit{substantially more reliably when guided by a structured interface}.
Under minimal prompting, performance is inconsistent and frequently fails to decompose compound requirements, whereas explicit syntactic constraints yield consistently higher accuracy across models.
Figure~\ref{fig:ap_extraction} summarizes performance across models and interfaces. Under the Minimal Interface, extraction quality varies substantially, with F1-scores ranging from 25.05\% (\texttt{GPT-3.5-turbo}) to 78.74\% (\path{Gemini-1.5-Flash}). Errors in this setting frequently involve semantic mismatches, incomplete phrase extraction, or failure to decompose compound statements. In contrast, the Detailed Interface yields consistently strong results across all models, with F1-scores exceeding 77\%, Jaccard similarity above 65\%, and Levenshtein similarity above 88\%, indicating substantially improved alignment with ground-truth propositions.

Explicit structural guidance substantially improves extraction accuracy. 
These results indicate that AP extraction strongly benefits from constraint-enforcing interfaces.
Under the Detailed Interface, catastrophic failures and over-generation are rare, while 
residual errors primarily involve semantic mismatch or incomplete decomposition of compound statements.
This suggests that, although structured prompting substantially mitigates extraction errors, it does not fully eliminate semantic ambiguity.
One validation criterion is adherence to the \emph{principle of maximum revelation}. For example, the sentence \emph{“Mary will not join the team and will not play the flute”} should be decomposed into $\neg p \land \neg q$, where $p$ denotes ``Mary will join the team'' and $q$ denotes ``Mary will play the flute.'' Several models, particularly under the Minimal Interface, treat the entire sentence as a single atomic proposition, obscuring its logical structure and violating this principle.
\begin{center}
\fbox{
\begin{minipage}{0.95\columnwidth}
\textbf{\rqtype{syn}{1} Summary.}
\emph{Atomic proposition extraction is unreliable under minimal prompting but
robust under Detailed interface.}
\end{minipage}
}
\end{center}

\noindent
\textbf{\RQtype{syn}{2}{How accurately can LLMs distinguish well-formed from malformed LTL formulas?}}

\begin{figure}[!thbp]
\vspace{2mm}

\centerline{\includegraphics[width=0.99\columnwidth]{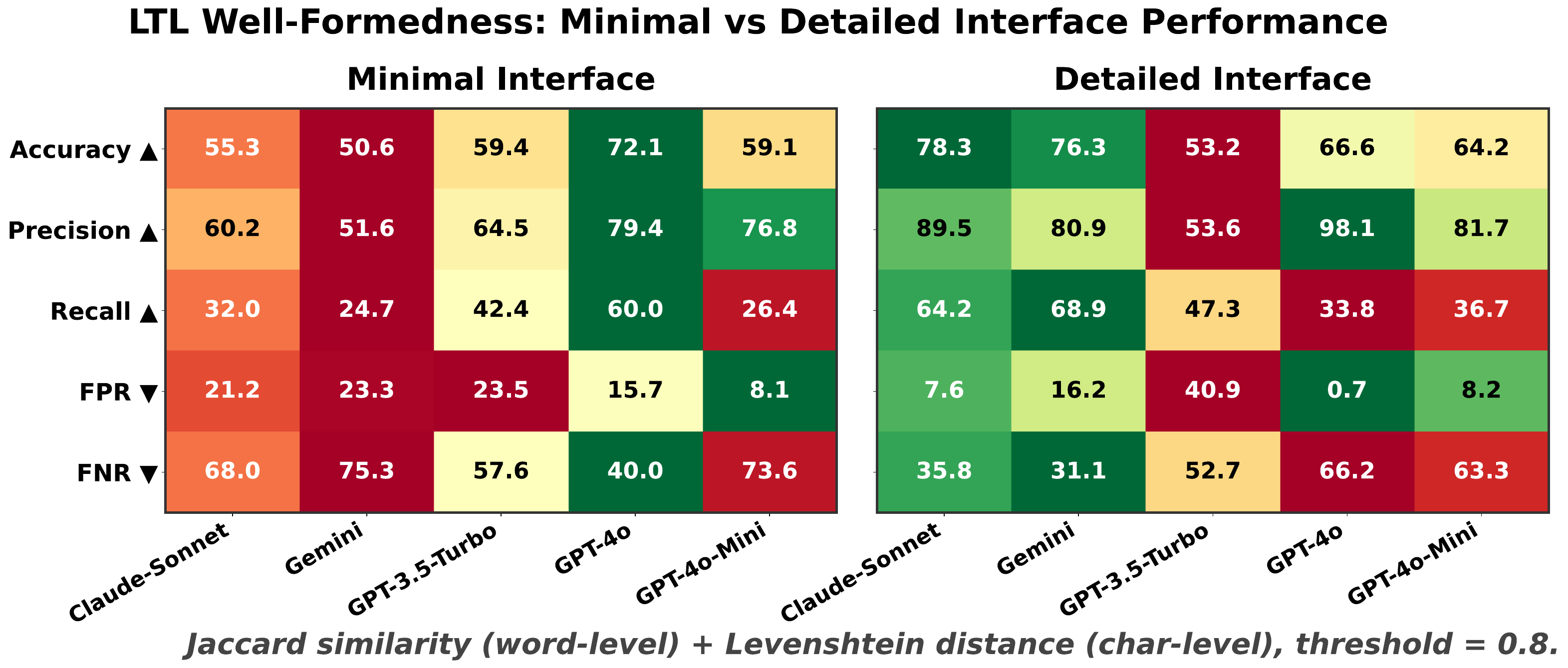}}

\Description{Heatmap of Detailed vs. Minimal Interface performance.}
\caption{Detailed vs. Minimal Interface performance. Accuracy represents the fraction of correctly classified formulas. Precision and Recall measure soundness and completeness, respectively. FPR and FNR denote the rates of incorrectly accepted malformed formulas and incorrectly rejected well-formed formulas.}
\label{fig:ltl_well_formedness}
\end{figure}

\noindent
\textbf{Answer.}
LLMs do \emph{not} consistently distinguish well-formed from malformed LTL formulas,
even when provided with explicit syntactic guidance.
Both soundness (rejecting malformed formulas) and completeness (accepting valid
formulas) are systematically violated across models and interfaces.
Figure~\ref{fig:ltl_well_formedness} summarizes classification performance across models and interfaces. Under the Detailed Interface, accuracy ranges from 53.22\% (\texttt{GPT-3.5-Turbo}) to 78.33\% (\texttt{Claude-Sonnet}); even in the best case, more than 20\% of formulas are misclassified. Performance further degrades under the Minimal Interface, where accuracy peaks at 72.13\% (\texttt{GPT-4o}) and several models approach chance-level behaviour. These results indicate that syntactic judgment remains unreliable even when explicit grammar information is provided. Notably, \texttt{GPT-3.5-Turbo} exhibits substantially lower distinguishing capability than all other evaluated models, a pattern that is consistent across datasets and evaluation metrics and suggests reduced robustness on tasks requiring precise temporal reasoning.

Malformed formulas incorrectly accepted (False positives) represent the most critical failure mode. Under the Detailed Interface, 14.71\% of malformed formulas are accepted, with false-positive rates ranging from 0.67\% (\texttt{GPT-4o}) to 40.89\% (\texttt{GPT-3.5-Turbo}). Under the Minimal Interface, this rate increases to 18.34\%. Conversely, false negatives are pervasive: 49.82\% of valid formulas are rejected under the Detailed Interface, increasing to 62.89\% under the Minimal Interface. 
Thus, models simultaneously fail to enforce syntactic soundness and to preserve completeness.

Errors reveal systematic weaknesses rather than isolated mistakes. Models frequently reject valid formulas involving nested or past-time operators (\texttt{Y}, \texttt{H}, \texttt{S}), and incorrectly penalize unconventional but legal atomic proposition names (\eg, rejecting \addformula{X(H(X(O(wykepokyro))))} because \addformula{wykepokyro} was deemed ``invalid''). At the same time, malformed formulas violating operator arity are sometimes accepted, including expressions where the binary \texttt{S} (Since) operator lacks a left operand. These inconsistencies indicate that exposure to grammar descriptions alone does not ensure reliable syntactic classification.

Because syntactic validity in formal verification is binary, these error rates limit the suitability of LLMs as standalone syntactic validators for LTL.
The observed error rates therefore preclude the use of LLMs as standalone
syntactic gatekeepers for LTL.
Any practical deployment must strictly encapsulate LLM outputs with formally
verified parsers and grammars.

\begin{center}
\fbox{
\begin{minipage}{0.95\columnwidth}
\textbf{\rqtype{syn}{2} Summary.}
\emph{LLMs cannot reliably judge LTL well-formedness.}
Explicit interfaces yield limited gains, but both soundness and completeness
remain violated across all models.
\end{minipage}
}
\end{center}

\noindent
\textbf{\RQtype{sem}{1}{Can LLMs generate LTL formulas that are semantically equivalent to the  ground truth?}}

\begin{figure}[!thbp]
\vspace{2mm}

\centerline{\includegraphics[width=0.99\columnwidth]{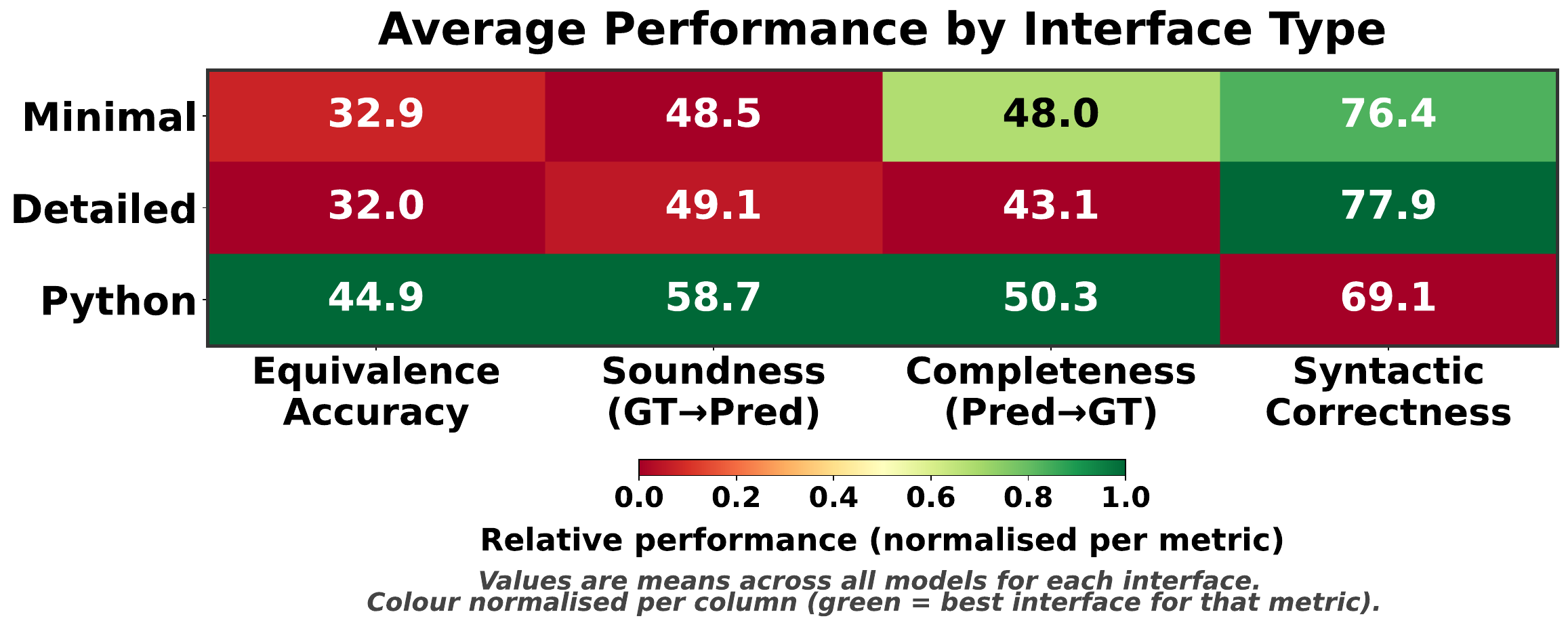}}

\Description{Heatmap of Detailed vs. Minimal vs. Python Interface performance of NL$\rightarrow$LTL translation.}
\caption{Semantic performance of NL$\rightarrow$LTL translation evaluated via bidirectional entailment. \textbf{Equiv.} denotes semantic equivalence ($\varphi_{\mathit{GT}} \Leftrightarrow \varphi_{\mathit{Pred}}$), \textbf{Sound.} denotes ground-truth-to-prediction entailment ($\varphi_{\mathit{GT}} \Rightarrow \varphi_{\mathit{Pred}}$), \textbf{Compl.} denotes prediction-to-ground-truth entailment ($\varphi_{\mathit{Pred}} \Rightarrow \varphi_{\mathit{GT}}$), and \textbf{Syntax} reports well-formed LTL outputs.}
\label{fig:rq1-main}
\end{figure}

\noindent
\textbf{Answer.}
LLMs \textit{rarely} generate LTL formulas that are semantically equivalent to the intended
ground truth.
Even when outputs are syntactically well-formed, exact semantic equivalence holds
for fewer than one quarter of cases, indicating a substantial gap between syntactic validity and semantic correctness.
Figure~\ref{fig:rq1-main} summarizes semantic performance across models and
interfaces. Exact semantic equivalence is rare: averaged across all settings,
only 24.37\% of predictions are logically equivalent to the ground truth, despite
nearly half of outputs being syntactically well-formed. Soundness (34.45\%)
slightly exceeds completeness (31.51\%), suggesting a tendency toward over-specification. These results show that syntactic validity alone is not a reliable indicator of semantic correctness.

Interface design appears to have a measurable but limited effect. AST-constrained Python
interfaces outperform text-based interfaces, improving equivalence accuracy to
over 61\% for the strongest models (\texttt{Gemini-2.5-Flash}, \texttt{Gemini-1.5-Pro}). However, even
under strict structural constraints, more than one-third of formulas remain
semantically incorrect. This indicates that enforcing syntactic well-formedness alone does not ensure correct temporal interpretation.
Stronger models exhibit both higher equivalence and more balanced soundness and
completeness, corresponding to more balanced handling of temporal scope and operator placement.
In contrast, weaker or interface-tuned models display pronounced
entailment asymmetries, frequently producing formulas that are either strictly
stronger or strictly weaker than the intended specification. Importantly, high
syntactic validity does not correlate with semantic correctness, reinforcing the
necessity of entailment-based evaluation.
While atomic proposition overlap is relatively high (57.2\%), operator agreement
drops to 36.7\%, and exact structural alignment falls to just 18.7\%. A primary failure pattern involves mis-scoping and incorrect nesting of temporal operators. These errors are invisible to
syntax-only checks but directly impact trace semantics.
Performance trends are consistent across the Textbook and Little Tricky Logic
datasets, with differences below 1 percentage point in Table~\ref{tab:rq1-dataset}.
This suggests that the observed errors are not strongly tied to dataset-specific phrasing.

\begin{table}[!thpb]
\vspace{2mm}

\centering
\caption{Semantic and syntactic correctness of predicted LTL formulas by dataset. (\textbf{Equiv.}) is the logical equivalence between the predicted and ground truth formulas ($\varphi_{\text{pred}} \equiv \varphi_{\text{gt}}$). (\textbf{Sound.}) holds when the ground truth formula entails the generated formula ($\varphi_{\text{gt}} \models \varphi_{\text{pred}}$), while (\textbf{Compl.}) holds when the predicted formula entails the predicted formula ($\varphi_{\text{pred}} \models \varphi_{\text{gt}}$). \textbf{Syntax} is the percentage of syntactically valid LTL formulas.}

\label{tab:rq1-dataset}
\begin{tabular}{lcccc}
\toprule
\textbf{Dataset} & \textbf{Equiv.} & \textbf{Sound.} & \textbf{Compl.} & \textbf{Syntax} \\
\midrule
Little Tricky Logic & 23.9\% & 33.7\% & 31.8\% & 48.7\% \\
Textbook            & 24.8\% & 35.2\% & 31.3\% & 50.0\% \\
\bottomrule
\end{tabular}
\end{table}

Our inspection of the incorrect outputs revealed the recurring issues while also distinguishing acceptable structural variations from true semantic errors. Models frequently misplace temporal operators, fail to preserve scope in nested expressions, or produce verbose representations for constraints.
For example, the input “\textit{No more than one thread can have the lock.}” is commonly expressed as $\neg(x1 \land x2)$, but the model instead produces a longer disjunction enumerating all exclusive cases. In other cases, phrases like ``\textit{after the last state in which x1 holds}'' are are frequently translated into formulas that alter temporal scope. 
Common issues encountered include but not limited to, operator precedence, the confusion of $G$ operator with $F$ where models occasionally replace $G$ with $F$ or vice versa, leading to altered semantics and spacing or parenthesis where minor syntax deviations are typically caught during parsing.

Overall, LLMs frequently produce syntactically valid Future-LTL formulas that do not preserve the intended semantics.
Bidirectional entailment exposes systematic over- and under-specification caused
by mis-scoped temporal operators.
These findings demonstrate that NL$\rightarrow$LTL translation cannot be reliably assessed without formal semantic checking.

\begin{center}
\fbox{
\begin{minipage}{0.95\columnwidth}
\textbf{\rqtype{sem}{1} Summary.}
\emph{LLMs rarely achieve semantic equivalence in NL$\rightarrow$LTL translation.}
Structured interfaces help, but mis-scoped temporal operators remain the dominant
failure mode.
\end{minipage}
}
\end{center}

\noindent
\textbf{\RQtype{sem}{2}{Can LLMs accurately assess the satisfiability of a trace against a given formula?}}

\noindent
\textbf{Answer.}
LLMs classify trace satisfiability with \textit{moderate and consistently improvable accuracy}, and performance improves systematically as structural information increases.
Across all evaluated models, Minimal interfaces yield near-chance to moderate performance, while Detailed and Python interfaces produce substantial and consistent gains.
As shown in Table~\ref{tab:trace-char-summary}, average F1 increases from approximately 57--60\% under Minimal interfaces to over 80\% under Python/AST interfaces.
\begin{table}[!thbp]
\centering
\small
\caption{Classification outcomes for satisfying and violating traces.
(\textbf{True Sat.}) denotes correctly accepted satisfying traces, while
(\textbf{False Sat.}) denotes satisfying traces incorrectly rejected.
(\textbf{True Viol.}) denotes correctly rejected violating traces, while
(\textbf{False Viol.}) denotes violating traces incorrectly accepted.
(\textbf{Sound.}) corresponds to acceptance accuracy on satisfying traces, and (\textbf{Compl.}) corresponds to rejection accuracy on violating traces. Results are aggregated by interface type.}

\label{tab:trace-char-summary}
\scalebox{0.8}{
\begin{tabular}{lcccccc}
\toprule
\textbf{Interface} &
\textbf{True Sat.} &
\textbf{False Sat.} &
\textbf{True Viol.} &
\textbf{False Viol.} &
\textbf{Sound.} &
\textbf{Compl.} \\
\midrule
Minimal  & 56.1\% & 43.9\% & 59.0\% & 41.0\% & 56.1\% & 59.0\% \\
Detailed & 64.3\% & 35.7\% & 71.2\% & 28.8\% & 64.3\% & 71.2\% \\
Python   & 78.0\% & 22.0\% & 81.3\% & 18.7\% & 78.0\% & 81.3\% \\
\bottomrule
\end{tabular}}
\end{table}
Performance exhibits a stable asymmetry: models more reliably identify violating traces than satisfying traces.
This asymmetry is reflected in higher completeness than soundness across interface types, indicating that models more consistently detect counterexamples than confirm global satisfaction conditions.
Because violating traces often depend on localized inconsistencies, whereas satisfying traces require validating temporal constraints across the full execution prefix, classification errors disproportionately arise when models must propagate constraints through nested temporal scopes.
Error inspection reveals that misclassifications concentrate in formulas containing multiple nested temporal operators, particularly repeated applications of $\mathbf{X}$ or mixed nesting of $\mathbf{F}$ and $\mathbf{G}$.
For example, $X(X(X(X(F(x_1)))))$
is frequently misclassified despite having straightforward satisfying traces, suggesting difficulty maintaining consistency across deeper temporal horizons.

In contrast, formulas containing single temporal operators (e.g., $F(x_1)$ or $G(x_1)$) are classified reliably across models and interfaces.
Accuracy degrades as operator depth increases, indicating that compositional temporal reasoning remains a limiting factor even when formulas are syntactically constrained.

Overall, results show that LLMs can perform non-trivial temporal trace evaluation, but reliability depends strongly on structured representations.
AST-based interfaces produce the most consistent improvements, reducing both false acceptance and rejection rates.

\begin{center}
\fbox{
\begin{minipage}{0.95\columnwidth}
\textbf{\rqtype{sem}{2} Summary.}
\emph{LLMs can classify trace satisfiability with moderate accuracy, improving substantially under structured interfaces.}
Errors concentrate in formulas with deeply nested temporal operators, a limitation in compositional temporal reasoning.
\end{minipage}
}
\end{center}

\noindent
\textbf{\RQtype{sem}{3}{Can LLMs generate valid traces for a given LTL formula?}}

\begin{figure}[!thbp]
\vspace{2mm}


\centerline{\includegraphics[width=0.99\columnwidth]{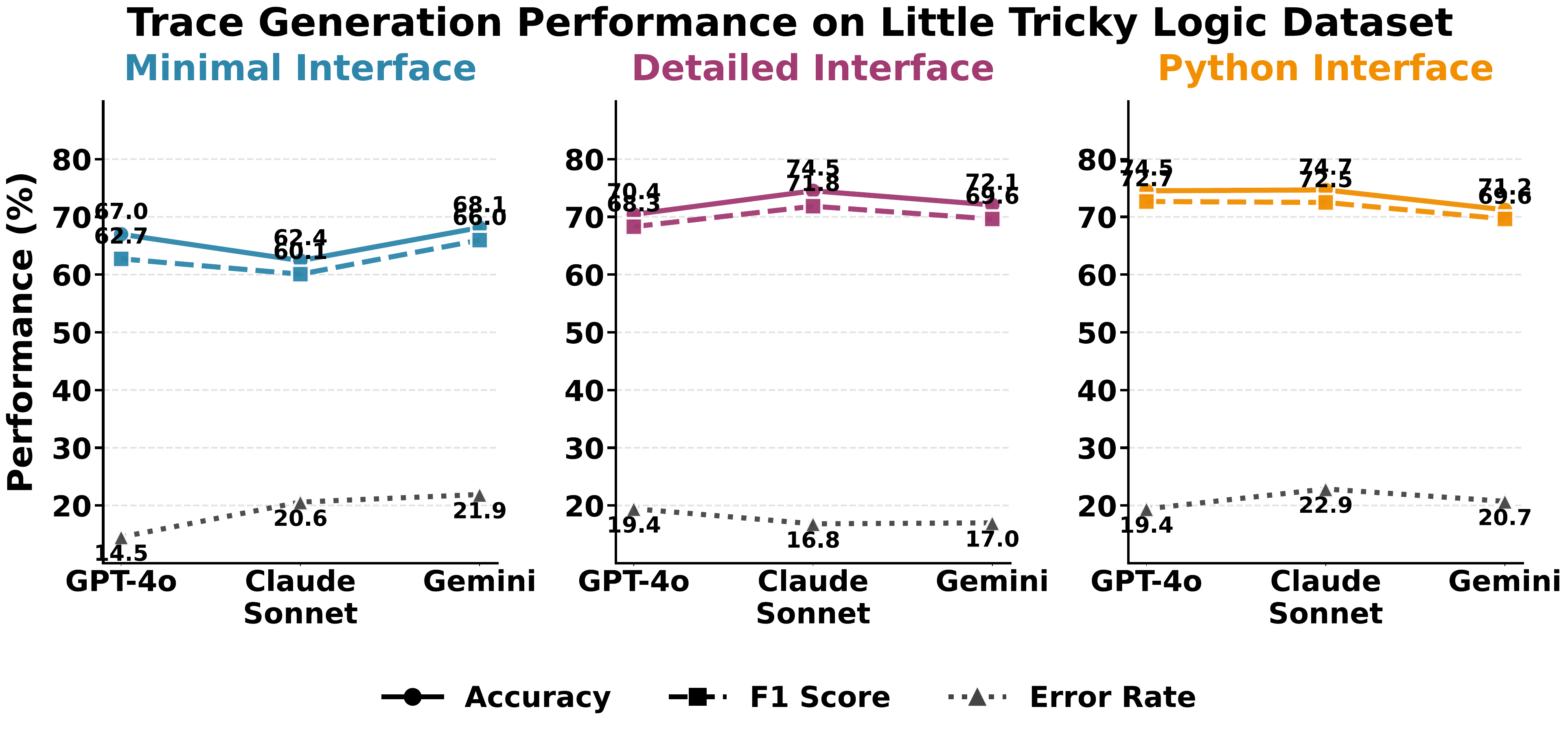}}

\Description{Line Chart of Detailed vs. Minimal vs. Python Interface performance of Trace generation.}
\caption{Trace generation results on Little Tricky Logic dataset \cite{Greenman_2022} for different interfaces. }
\label{fig:trace-gen}
\end{figure}

\noindent
\textbf{Answer.}
LLMs generate traces that satisfy or violate LTL formulas with \textit{moderate reliability}, achieving F1 scores generally between 55\% and 70\% across models and interfaces.
Performance improves under structured prompting, but remains less stable than trace classification.
Figure~\ref{fig:trace-gen} summarizes performance across interfaces.

Generating satisfying traces is consistently easier than constructing violating traces.
Producing a falsifying trace requires identifying a counterexample that breaks a global temporal condition without accidentally satisfying the specification.
This requirement becomes increasingly brittle as operator nesting depth increases, particularly for formulas combining $\mathbf{F}$, $\mathbf{G}$, and repeated $\mathbf{X}$ operators.

Failures frequently occur when traces must maintain invariants over multiple time steps.
For example, formulas containing nested next operators require propagating satisfaction constraints across several future states, increasing the likelihood of local inconsistencies that invalidate the trace.
Compared to NL$\rightarrow$LTL, trace generation demonstrates higher overall reliability, suggesting that models more readily construct concrete executions than derive symbolic temporal specifications.
However, generated traces often exhibit small logical inconsistencies that prevent full semantic alignment with the formula, indicating incomplete internalization of temporal execution semantics.
Structured Python/AST interfaces produce the strongest average results, but also introduce higher variance across formulas.
This suggests that structural constraints help guide generation, but do not eliminate sensitivity to formula complexity.

Overall, LLMs can produce candidate traces consistent with temporal specifications, but these traces cannot be assumed correct without formal validation.
Trace generation should therefore be interpreted as producing plausible executions.

\begin{center}
\vspace{0.2cm}
\fbox{
\begin{minipage}{0.95\columnwidth}
\textbf{\rqtype{sem}{3} Summary.}
LLMs can generate traces consistent with LTL formulas, but reliability remains moderate.
Performance decreases as temporal nesting depth increases.
Generated traces should be formally verified before use.
\end{minipage}
}
\end{center}

\begin{table*}[!hptb]
\centering
\caption{\textbf{Summary of Key Findings}}
\label{tab:summary}
\resizebox{\textwidth}{!}
{%
\begin{tabular}{|c|l|c|c|c|}
\hline
\textbf{RQ} & \textbf{Task} & \textbf{$\mathcal{M}$ (Minimal Interface)} & \textbf{$\mathcal{D}$ (Detailed Interface)} & \textbf{$\mathcal{P}$ (Python Interface)} \\
\hline
\hline
$\text{RQ}_{\text{syn}}^1$ & Atomic Proposition Extraction & $J=\textbf{0.83}$ (\claudesonnet) & $J=\textbf{0.86}$ (+3.61\%) (\gemini) & N/A \\
& & $L_d=\textbf{3.66}$ (\claudesonnet) & $L_d=\textbf{3.05}$ (-16.67\%) (\gemini) & \\
\hline
$\text{RQ}_{\text{syn}}^2$ & LTL Well-Formedness Validation & Acc=\textbf{52.19\%} (Avg.) & Acc=\textbf{84.33\%} ($\Delta_{\text{acc}}=+10.86\%$) (\gemini) & N/A \\
& & $F_1=\textbf{57.96\%}$ (Avg.) & $F_1=\textbf{85.08\%}$ ($\Delta_{F_1}=+17.87\%$) (\gemini) & \\
& & & $\epsilon_r=\textbf{20.4\%}$ (Total Reduction) & \\
\hline
$\text{RQ}_{\text{sem}}^1$ & NL$\to$Future LTL Generation & $\text{Equiv}=\textbf{65.56\%}$ (\gemini) & $\text{Equiv}=\textbf{64.75\%}$ (\claudesonnet) & $\text{Equiv}=\textbf{72.13\%}$ (\claudesonnet) \\
& & $F_1=\textbf{72.27\%}$ (\gemini) & $F_1=\textbf{72.35\%}$ (\claudesonnet) & $F_1=\textbf{77.80\%}$ (\claudesonnet) \\
& & & & $\phi_{\text{GT}} \models \phi_{\text{pred}}$: \textbf{+23.8\%} (vs $\mathcal{M}$, Stat. Sig.) \\
\hline
$\text{RQ}_{\text{sem}}^2$ & Trace Satisfiability Classification & Acc=\textbf{76.00\%} (\gemini) & Acc=\textbf{80.72\%} (\claudesonnet) & Acc=\textbf{100.00\%} (\geminitwofiveflash) \\
& & & $\Delta_{\text{acc}}=+7.7\%$ (vs $\mathcal{M}$) & $\Delta_{\text{acc}}=+\textbf{22.4\%}$ (vs $\mathcal{M}$, Stat. Sig.) \\
\hline
$\text{RQ}_{\text{sem}}^3$ & Trace Generation & Acc=\textbf{68.14\%} (\gemini) & Acc=\textbf{74.51\%} (\claudesonnet) & Acc=\textbf{74.67\%} (\geminitwofiveflash) \\
& & & (Claude \fewshot) & ($<20\%$ for \fewshot) \\
\hline
$\text{RQ}_{\text{sem}}^4$ & NL$\to$Past LTL Generation & $\text{Equiv}=\textbf{65.56\%}$ (\gemini) & $\text{Equiv}=\textbf{69.96\%}$ (\gemini) & $\text{Equiv}=\textbf{72.96\%}$ (\geminitwofiveflash) \\
& & $F_1=\textbf{72.27\%}$ (\gemini) & $F_1=\textbf{76.04\%}$ (\gemini) & $F_1=\textbf{79.23\%}$ (\geminitwofiveflash) \\
& & & $\text{Syn}_{\text{corr}}=\textbf{99.65\%}$ (\claudesonnet) & $\text{Syn}_{\text{corr}}=\textbf{96.84\%}$ (\geminitwofiveflash) \\
\hline
\end{tabular}
}

\vspace{0.2cm}
\footnotesize
$\mathcal{M} = $ \minpro; $\mathcal{D} = $ \detpro; $\mathcal{P} = $ \pypro; $J = $ Jaccard Similarity; $L_d = $ Levenshtein Distance; $\Delta = $ Absolute Change; $\epsilon_r = $ Error Reduction; $\text{Acc} = $ Accuracy; $\text{Equiv} = $ Equivalence Accuracy; $F_1 = $ F1-Score; $\text{Syn}_{\text{corr}} = $ Syntactic Correctness Rate; $\phi_{\text{GT}} \models \phi_{\text{pred}} = $ Ground Truth entails Prediction (Semantic Completeness); N/A = Not Applicable; Stat. Sig. = Statistically Significant.
Performance numbers represent the \textbf{best observed result} for a model within that specific strategy and task, unless explicitly stated as an average or a range. 
\end{table*}

\noindent
\textbf{\RQtype{sem}{4}{Can LLMs generate semantically equivalent \textbf{Past}-LTL formulas from natural language descriptions?}}

\begin{figure}[!thbp]
\vspace{2mm}

\centerline{\includegraphics[width=0.99\columnwidth]{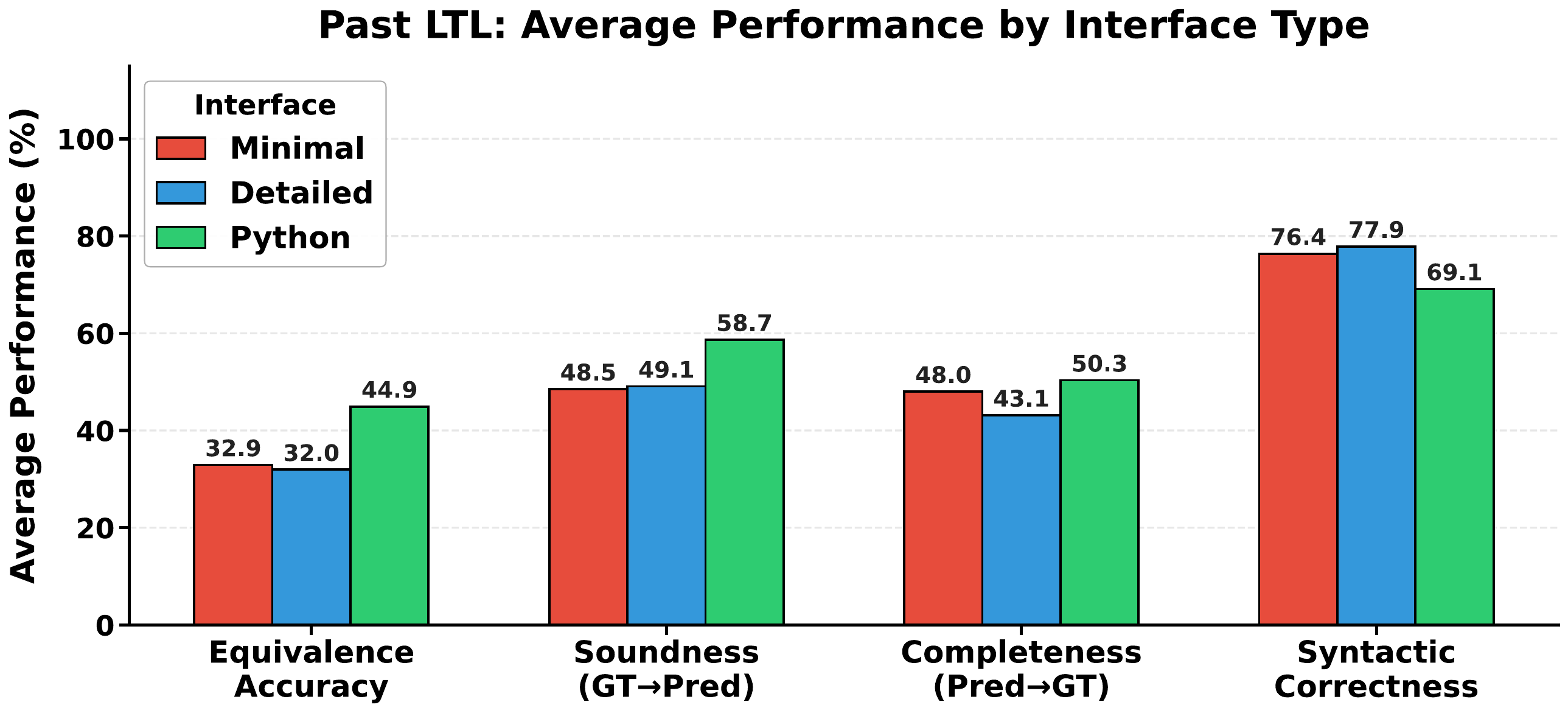}}

\Description{Bar chart comparing Detailed vs. Minimal vs. Python interface performance of NL$\rightarrow$LTL (Past) translation evaluated via bidirectional entailment and semantic equivalence.}

\caption{Semantic performance of NL$\rightarrow$LTL (Past) translation evaluated via bidirectional entailment. \textbf{Equiv.} denotes semantic equivalence ($\varphi_{\mathit{GT}} \Leftrightarrow \varphi_{\mathit{Pred}}$), \textbf{Sound.} denotes ground-truth-to-prediction entailment ($\varphi_{\mathit{GT}} \Rightarrow \varphi_{\mathit{Pred}}$), \textbf{Compl.} denotes prediction-to-ground-truth entailment ($\varphi_{\mathit{Pred}} \Rightarrow \varphi_{\mathit{GT}}$), and \textbf{Syntax} reports well-formed LTL outputs.}
\label{fig:past-ltl-summary}
\end{figure}

\noindent
\textbf{Answer.}
LLMs can generate semantically meaningful Past-LTL formulas, but performance
remains sensitive to structural constraints.
Under AST-constrained Python interfaces, top-performing models achieve
69--72\% semantic equivalence, compared to 59--63\% under text-based interfaces.
Soundness and completeness remain relatively balanced, indicating that most
errors arise from subtle structural mismatches rather than systematic bias
toward over- or under-specification.
Figure~\ref{fig:past-ltl-summary} summarizes semantic performance across interfaces.
Text-based interfaces exhibit high variance across models, often producing
syntactically valid formulas that are not semantically equivalent.
Structured interfaces reduce this variance and improve equivalence rates,
suggesting that constraining operator composition supports more stable temporal reasoning across logic fragments.
Error inspection shows that failures primarily involve incorrect operator
scoping and precedence interactions between past operators such as $Y$, $S$, and
implication.
For example, the sentence \emph{``It previously held that if $x_1$ was true then $x_2$ was true before that''}
often produces formulas with incorrect nesting between $Y$ and $S$,
altering temporal dependencies despite syntactic validity.
Models also occasionally confuse related past operators (\eg, $Y$ vs.\ $H$)
or substitute equality ($=$) for logical equivalence ($\leftrightarrow$),
yielding formulas that parse correctly but fail semantic checks.

Compared to Future-LTL generation (RQ$_{\text{sem}}^{1}$),
Past-LTL translation follows similar performance ordering across models but
exhibits slightly lower equivalence overall, indicating that additional operator
interactions introduce further semantic drift.

\begin{center}
\fbox{
\begin{minipage}{0.95\columnwidth}
\textbf{\rqtype{sem}{4} Summary.}
LLMs can generate Past-LTL formulas with moderate semantic accuracy.
Structural constraints improve consistency, but operator scoping errors remain common.
\end{minipage}
}
\end{center}

\noindent
Across semantic tasks, performance rankings remain broadly consistent:
models that perform well on NL$\rightarrow$LTL generation (RQ$_{\text{sem}}^{1}$)
also perform well on trace classification and generation (RQ$_{\text{sem}}^{2,3}$).
However, trace-based tasks generally exhibit higher accuracy than formula
synthesis, suggesting that recognizing temporal behavior is easier than
constructing semantically equivalent symbolic specifications.
Past-LTL translation (RQ$_{\text{sem}}^{4}$) follows the same ordering with
slightly lower equivalence rates overall.

\noindent
Table~\ref{tab:summary} summarizes trends identified in our experimental results.

In-depth results of the \rqtype{}{}s can be found in the Appendix~\ref{sec:appendix:results}.

\section{Evaluation with Security Requirements}
\label{sec:security}

We evaluate whether trends observed in the main benchmark generalize to
security-oriented specifications, and whether providing atomic proposition (\textbf{\textit{AP}})
mappings impacts end-to-end NL$\rightarrow$Future LTL translation performance.
Security requirements often encode implicit dependencies between authorization,
policy enforcement, and system state, making correct temporal scoping essential.

\subsection{Evaluation Setup}
\noindent\textbf{Model.}
The main benchmark experiments were originally conducted using
Claude-3.5-Sonnet, which has since been deprecated.
We therefore use its successor, Claude-4.5, for all experiments in this section.
To ensure fair comparison across domains, we construct a matched baseline by
randomly sampling 56 requirements from \dstype{tricky}{1} and rerunning the
NL$\rightarrow$Future LTL experiment using identical prompting configurations.

\noindent\textbf{Dataset and experimental conditions.}
We randomly selected 56 ground truth entries (\ie \dstype{verify}{7}) 
from the VERIFY dataset~\cite{quansah2026verify} covering authentication persistence, session validity, re-authentication triggers, and policy-driven access control. NL descriptions were minimally normalized to preserve logical structure while removing stylistic redundancy, ensuring evaluation isolates temporal reasoning without altering operator nesting or temporal depth.
We select 56 samples of baseline dataset (\dstype{tricky}{1}) to match sample size, 
and to enable manual verification of atomic proposition alignment in the NL-only condition.
Two settings are considered:

\begin{itemize}
    \item \textbf{NL+AP:} ground-truth atomic propositions are provided
    \item \textbf{NL-only:} propositions must be inferred and are manually aligned to ground truth prior to semantic evaluation
\end{itemize}

All three prompting strategies from the main benchmark are evaluated:
zero-shot, few-shot, and zero-shot self-refinement.

\subsection{Results}

\textbf{Security domain results.}
Table~\ref{tab:security-combined} shows results on \dstype{verify}{7}
when AP mappings are provided.
Performance is comparable to the baseline, with self-refinement achieving
the highest equivalence (59.38\%) and F1 (60.9\%).
For No AP experiment, performance decreases across all prompting strategies, with F1 ranging
from 45.0\% to 53.3\%.
The higher number of Not Meaningful cases indicates that proposition
misalignment has greater impact in the security domain.

\begin{table}[t]
\centering
\caption{Security dataset results: Comparing AP  vs. No AP + manual alignment.}
\label{tab:security-combined}
\scalebox{0.75}{
\begin{tabular}{llcccccc}
\toprule
\textbf{Condition} & \textbf{Approach} & \textbf{Equiv.} & \textbf{Sound.} & \textbf{Compl.} & \textbf{Prec.} & \textbf{Recall} & \textbf{F1} \\
\midrule
\multirow{3}{*}{With AP} & Few-shot & 37.50 & 46.88 & 58.82 & 37.50 & 46.88 & 41.67 \\
 & Zero-shot & 53.12 & 65.62 & 64.71 & 53.12 & 65.62 & 58.72 \\
 & Self-refine & 59.38 & 62.50 & 70.59 & 59.38 & 62.50 & 60.90 \\
\cmidrule(r){1-8}
\multirow{3}{*}{No AP} & Few-shot & 44.44 & 61.11 & 46.51 & 44.44 & 61.11 & 51.46 \\
 & Zero-shot & 47.22 & 61.11 & 44.19 & 47.22 & 61.11 & 53.28 \\
 & Self-refine & 40.00 & 51.43 & 42.86 & 40.00 & 51.43 & 45.00 \\
\bottomrule
\end{tabular}
}
\end{table}

\noindent
\textbf{Matched baseline (Little Tricky Logic).}
Table~\ref{tab:baseline-56} shows results on the 56-sample baseline.
Providing AP mappings yields equivalence between 46.3\% and 53.9\%.
Without AP mappings, equivalence remains similar (48.1--49.0\%),
indicating that proposition identification is not a major bottleneck for
general-domain specifications.

\noindent
\textbf{Discussion.}
Results indicate that the security domain introduces modest additional
difficulty compared to the matched baseline.
While performance with AP mappings is similar across domains,
the performance drop in the NL-only condition is larger for security
requirements.
This suggests that correct identification of atomic propositions plays a
greater role when specifications encode domain-specific dependencies
between authentication events, policy checks, and system responses.
\begin{table}[t]
\centering
\caption{Matched baseline results from \dstype{tricky}{1} on 56 samples}
\label{tab:baseline-56}
\scalebox{0.75}{
\begin{tabular}{llcccccc}
\toprule
\textbf{Condition} & \textbf{Approach} & \textbf{Equiv.} & \textbf{Sound.} & \textbf{Compl.} & \textbf{Prec.} & \textbf{Recall} & \textbf{F1} \\
\midrule
\multirow{3}{*}{With AP}
& Few-shot & 53.85 & 71.15 & 67.31 & 53.85 & 71.15 & 61.30 \\
& Zero-shot & 46.30 & 72.22 & 53.70 & 46.30 & 72.22 & 56.42 \\
& Self-refine & 50.94 & 77.36 & 58.49 & 50.94 & 77.36 & 61.43 \\
\midrule
\multirow{3}{*}{No AP}
& Few-shot & 48.08 & 65.38 & 55.77 & 48.08 & 65.38 & 55.41 \\
& Zero-shot & 48.08 & 69.23 & 57.69 & 48.08 & 69.23 & 56.75 \\
& Self-refine & 49.02 & 70.59 & 60.78 & 49.02 & 70.59 & 57.86 \\
\bottomrule
\end{tabular}
}
\end{table}
Across both domains, the dominant source of error remains incorrect temporal
scoping rather than syntactic invalidity, confirming that temporal reasoning
continues to be the primary challenge in NL$\rightarrow$LTL translation.

\begin{center}
\fbox{
\begin{minipage}{0.95\columnwidth}
\textbf{Summary.}
The security domain introduces measurable additional difficulty
(3--16 point equivalence drop vs.\ \dstype{tricky}{1} baseline).
AP grounding is substantially more impactful in the security domain
than in general benchmarks (6--19 vs.\ 1--2 point gap).
Temporal operator mis-scoping remains the dominant failure mode
across all conditions, confirming that domain complexity amplifies
existing structural reasoning limitations.
\end{minipage}
}
\end{center}

\section{Related Work}
This section presents relevant existing efforts. 

\mypara{Natural Language to Temporal Logic.} Translation of NL requirements into formal temporal specifications has been studied extensively. Early approaches relied on property specification patterns~\cite{10.1145/302405.302672}, controlled natural languages~\cite{10.1007/11901433_41,Farrell_2024}, and syntax-directed translation~\cite{10.1007/978-3-540-30206-3_12,10.5555/332656}. While effective within restricted domains, these techniques depend on hand-crafted grammars or templates, limiting scalability and requiring substantial manual effort. Evaluation is typically pattern- or syntax-driven, without assessing whether generated specifications preserve intended behavior over time.

\mypara{Security and Privacy Specifications.} The VERIFY dataset~\cite{quansah2026verify} introduces a multi-domain collection spanning authentication, access control, and privacy policies, constructed via an intermediate representation supporting provable translation to LTL. It emphasizes 
the importance of correctly identifying atomic propositions representing system states and information flows. Complementary work has examined temporal logic in security policy verification under realistic threat models~\cite{10917671}. Unlike datasets derived primarily from textbooks or synthetic benchmarks~\cite{zhang2024fusionlargelanguagemodels,10.1007/978-3-031-37703-7_18,Fuggitti_Chakraborti_2024}, these efforts capture domain semantics closer to real-world verification. Our evaluation incorporates a VERIFY subset to assess whether semantic correctness trends transfer to security-relevant specifications.

\mypara{LLMs for Code Generation.} Recent work has investigated LLMs for generating logical formulas and program code~\cite{chen2021evaluating,wang2021codet5,li2023starcoder}, primarily reporting success in syntactic validity or string-level similarity. Few-shot learning and chain-of-thought reasoning have been proposed to improve logical coherence~\cite{10.5555/3600270.3602070}, yet evaluation is rarely grounded in formal semantic correctness, meaning models may produce well-formed outputs while failing at intended behavior.

\mypara{Syntactic vs.\ Semantic Correctness.} A foundational distinction in formal methods separates syntactic correctness (whether a formula is well-formed) from semantic correctness (whether it correctly characterizes system behavior over execution traces)~\cite{Pnueli_1977,baier2008principles}. Syntactic validity provides no guarantee of semantic adequacy~\cite{10.1145/302405.302672}, yet most \nlltltask and LLM-based approaches emphasize the syntactic, substantially overestimating reliability in verification contexts. Our work distinctively evaluates both dimensions independently.

\section{Discussion and Recommendations}

This section highlights practical implications of our evaluation for
secure system specification and verification workflows.

\mypara{Data contamination.}
We cannot guarantee benchmark data did not appear in model training corpora,
particularly given public GitHub availability.
If contamination exists, results should be interpreted as an upper bound on
LLM capability.

\mypara{Requirements to logic.}
Our evaluation focuses on short NL fragments, while real-world security
requirements are typically longer and more interdependent.
Results therefore represent a lower bound on end-to-end NL-to-formal-specification
performance, with scaling subject to token limits and long-context reasoning
constraints.

\mypara{Impact of data regularity.}
Semantic equivalence is consistently higher on \dstype{book}{6} than on
\dstype{tricky}{1}, whose edge-case constructions expose structural weaknesses
in NL$\rightarrow$LTL translation.
Errors arise primarily from incorrect temporal scoping rather than failure to
identify relevant domain concepts, indicating that reasoning about event ordering
remains the principal challenge.

\mypara{Security Implications.} Security requirements rely on implicit assumptions regarding system state and trust. 
Our results on \dstype{verify}{7}
show that the overall performance 
primarily depends on the reliable grounding of atomic propositions. Inconsistent symbolic representations and incorrect temporal scoping risk mis-specified policies that appear valid but fail to enforce security guarantees. Thus, \nlltltask must be viewed as a decision-support tool and not a fully automated process, necessitating human oversight and stable proposition sets to ensure verifiable security.

\mypara{Prompting impacts.}
Prompt design substantially affects performance.
Few-shot prompting generally outperforms zero-shot variants, while
self-refinement yields moderate improvement.
Detailed prompts improve performance across tasks, and reformulating the task
as Python code completion consistently produces the strongest results,
suggesting that structured output constraints improve reliability for
security-critical specification generation.

\section{Threats to Validity}
\label{sec:threats}
We discuss threats to validity following empirical methodology,
distinguishing between internal, external, and construct validity.

\mypara{Internal Validity.} Our results depend on the correctness of the evaluation pipeline and NuSMV. While NuSMV is well-established, implementation errors in trace encoding or formula translation could affect outcomes. We mitigate this through systematic validation and uniform application across all models and interfaces.

\mypara{External Validity.} The evaluation focuses on short, self-contained NL requirements, whereas real-world specifications are often longer, more contextual, and cross-referential. Results should therefore be interpreted as lower bounds on the challenges posed by industrial requirements. Model performance may also shift as LLMs are updated, limiting strict reproducibility over time.

\mypara{Model availability over time.} This project spanned an extended period during which LLM APIs and model versions evolved. We report results using the best-performing available configuration per interface at the time of evaluation. Although exact model versions may not remain accessible, relative trends across interfaces and prompting strategies are consistent.

\mypara{Construct Validity.} LLMs may have been exposed to portions of the evaluation data during training, particularly for publicly available datasets. This does not invalidate comparative findings across interfaces, tasks, or evaluation criteria; if contamination exists, results represent an upper bound on performance under favorable conditions.

\mypara{Availability of Security-relevant Datasets.}
Although the \dstype{verify}{7} dataset~\cite{quansah2026verify} is large-scale (200k+ specifications across multiple domains), we select only a subset for \nlltltask evaluation with manual proposition verification for No-AP task.
As a result, our security-domain analysis is limited to a smaller curated sample, which may not fully represent the diversity of real-world security policies.
Future work would benefit from broader benchmarks explicitly targeting authentication, authorization, and compliance constraints.

\section{Conclusion}

This paper presents a systematic evaluation of existing LLMs’ efficacy for the \nlltltask task, motivated by a practical question for
the security and privacy community: whether LLMs can reliably
serve as natural-language front ends to testing, monitoring, policy analysis, and verification tools that take LTL formulas as input. Our
evaluation suggests that the answer is not yet. Although LLMs
often produce syntactically plausible formulas, their performance
on the semantic aspects of LTL remains substantially weaker. In security- and privacy-relevant settings, even small semantic deviations can change the meaning of authentication, authorization,
policy-enforcement, and monitoring requirements, causing downstream tools to analyze or enforce the wrong property. More generally, our results suggest that using LLMs for temporal semantic
reasoning tasks such as model checking is not yet well motivated,
and that such tasks should remain with dedicated symbolic reasoners. One of our surprising findings was that posing \nlltltask
as a Python code-completion or comprehension task substantially
improves LLM performance, though it does not eliminate the reliability gap. To enable fair, large-scale evaluation and comparison
with prior efforts, future fine-tuning efforts for the \nlltltask task
should accept NL-to-proposition mappings as input, and should
keep testing data private to reduce data contamination.





\begin{acks}
We thank the anonymous reviewers for their suggestions and comments. 
This material is based on research that is in part supported by the State University of New York's Empire Innovation Program and DARPA (SciFy program) under agreement number HR00112520301. The U.S. Government is authorized to reproduce and distribute reprints for Governmental purposes notwithstanding any copyright notation thereon. The views and conclusions contained herein are those of the authors and should not be interpreted as necessarily representing the official policies or endorsements, either express or implied, of DARPA, the State of New York, or the U.S. Government.
\end{acks}

\bibliographystyle{ACM-Reference-Format}
\bibliography{sample}

\appendix
\newpage
\section*{Appendix}

\section{Prompting Templates}
\label{appendix:prompts}
This appendix provides full prompt templates used for the \minpro, \detpro, and \pypro prompting strategies. We focus on the prompt used in translating the \texttt{NL} into the future \ltl.

\noindent
\textbf{Base Prompt for NL-to-Future LTL under \minpro strategy \label{appendix:minimalprompt}}
\begin{lstlisting}[language=Python]
def generate_base_prompt(natural_language, atomic_propositions):
    """Generate the base prompt for any approach."""
    return (
        "You are a Linear Temporal Logic (LTL) Parser. Your task is to convert a given "
        "Natural Language statement to an LTL formula, using the provided mapping of "
        "natural language phrases to atomic propositions.\n\n"
        "LTL Symbols:\n"
        "- AND: &\n"
        "- OR: |\n"
        "- NOT: !\n"
        "- IMPLIES: ->\n"
        "- BIIMPLICATION: <->\n"
        "- NEXT: X\n"
        "- EVENTUALLY: F\n"
        "- ALWAYS: G\n"
        "- UNTIL: U\n\n"
        f"Natural Language statement: {natural_language}\n"
        f"Atomic Propositions mapping: {atomic_propositions}\n\n"
    )
\end{lstlisting}

\noindent
\textbf{Base Prompt for NL-to-Future LTL under \detpro strategy \label{appendix:detailedprompt}}


\begin{lstlisting}[language=Python]
def generate_base_prompt(natural_language, atomic_propositions):
    """Generate the base system prompt for LTL conversion task"""
    return f"""You are a teacher who is proficient in propositional linear temporal logic (LTL).
In propositional linear temporal logic, you have three elements:
- propositional variables;
- logical connectives/operators;
- linear temporal connectives/operators.
Logical connectives/operators in propositional linear temporal logic include:
logical AND or conjunction (represented as &),
logical OR or disjunction (represented as |),
logical Not or negation (represented as !),
logical implication or entailment (represented as ->),
logical equivalence or bi-implication (represented as <->).
Linear temporal logic connectives/operators in propositional linear temporal logic include:
Next or tomorrow (represented as X),
Eventually or future (represented as F),
Globally or henceforth (represented as G),
Until (represented as U)
Yesterday or last (represented as Y),
Once (represented as O),
Historically (represented as H),
Since (represented as S)
Here is a BNF grammar for the syntax of propositional linear temporal logic formulas.
<formula> ::= <ap> | <TRUE> | <FALSE> | <formula> "&" <formula> | <formula> "|" <formula> |
"!" <formula> | "(" <formula> ")" | <formula> "->" <formula> | <formula> "<->" <formula> 
| <formula> "U" <formula> | "F" <formula> | "G" <formula> | "X" <formula> |
<formula> "S" <formula> | "O" <formula> | "H" <formula> | "Y" <formula>
<TRUE> ::= "true" | "True" | "TRUE" 
<FALSE> ::= "false" | "False" | "FALSE"
<ap> ::= It is the set of propositional logic variables which should start with any letters (small or capital) followed by an alphanumeric string.
Simply put, <ap> ::= [a-zA-Z][a-zA-Z0-9]* in regular expression.
The semantics of a propositional linear temporal logic formula are defined with respect to a linear trace \sigma and a position i in the trace. The position i is a non-negative number (i.e., 0 or any positive whole number).
Each element of a trace is a substitution (a mapping from propositional variables to either true or false).
In short, a trace \sigma is a sequence of states, where each state assigns truth values to propositional variables.
Given a trace \sigma, a position i in \sigma where the temporal logic formula is being evaluated, we can have the following semantics:
(1) \sigma and i always satisfy true
(2) \sigma and i falsify false
(3) In \sigma and i, a proposition a is true if and only if the current substitution \sigma[i] satisfies a
(4) In \sigma and i, p & q is true if and only if both p and q are true in \sigma and i
(5) In \sigma and i, p | q is true if and only if one of p or q are true in \sigma and i
(6) In \sigma and i, !p is true if and only if p is false in \sigma and i
(7) In \sigma and i, p -> q is true if and only if when p is true then q is true in \sigma and i
(8) In \sigma and i, p <-> q is true if and only if both p and q are both true or p and q are both false in \sigma and i
(9) In \sigma and i, X p is true if and only if p is true in \sigma at position i + 1
(10) In \sigma and i, Y p is true if and only if i > 0 and p is true in \sigma at position i - 1
(11) In \sigma and i, O p is true if and only if p is true in \sigma at position i or in \sigma at any position lower than i
(12) In \sigma and i, F p is true if and only if p is true in \sigma at position i or in \sigma at any position greater than i
(13) In \sigma and i, H p is true if and only if p is true in \sigma at position i and in \sigma at all positions lower than i
(14) In \sigma and i, G p is true if and only if p is true in \sigma at position i and in \sigma at all positions greater than i
(15) In \sigma and i, p S q is true if and only if q is true in \sigma at position i and p is true in \sigma at all positions from some position j <= i to position i
We say a trace \sigma satisfies a propositional LTL formula f if and only if \sigma satisfies f in the 0th position of \sigma.
You are now trying to convert a natural language English text into a propositional linear temporal logic formula. The input for this task will have two parts: natural language English sentences followed by natural language to propositional variable mapping. The input mapping from propositional variables to the natural language fragment having the form (variable_name -> "English sentence fragment") dictates the meaning of the propositional variable. For this task, you can only use the propositional variable given to you as part of the task input. Do not introduce any new propositional variables other than what is given to you. When generating the answer for the given natural language text to convert to LTL, stop providing additional explanations. Your output should only contain the formula in a single line.
Now convert the following:
Natural Language: {natural_language}
Proposition Mapping: {atomic_propositions}
When converting the natural language sentences in English to LTL, you cannot use any of the past temporal operators. The LTL formula you should output can only contain the following operators/connectives: &, |, !, ->, <->, X, F, U, G, Y, O, H, S
Very Important Syntax Rules
You must use only the following symbols in your formula:
- & for logical AND
- | for logical OR
- ! for logical NOT
- -> for implication
- <-> for bi-implication
- F, G, X, U for temporal operators
Your output must be a **single-line formula** using only the allowed syntax above, and must strictly follow the BNF grammar provided.
Convert the natural language to LTL formula:
"""
\end{lstlisting}


\noindent
\textbf{Base Prompt for NL-to-Future LTL under \pypro strategy \label{appendix:pythonprompt}}


\begin{lstlisting}[language=Python]
def generate_base_prompt(natural_language, atomic_propositions):
    return f"""
You are a teacher who is proficient in propositional linear temporal logic (LTL) and Python. You are given the following Python class structure that defines how LTL formulas should be represented:
```python
from dataclasses import dataclass
from typing import *
class Formula:
    pass
@dataclass
class AtomicProposition(Formula):
    name : str
@dataclass
class Literal(Formula):
    name : str
@dataclass
class LNot(Formula):
    Formula: Formula
@dataclass
class LAnd(Formula):
    left: Formula
    right: Formula
@dataclass
class LOr(Formula):
    left: Formula
    right: Formula
@dataclass
class LImplies(Formula):
    left: Formula
    right: Formula
@dataclass
class LEquiv(Formula):
    left: Formula
    right: Formula    
@dataclass
class Since(Formula):
    a : Formula
    b : Formula
@dataclass
class Until(Formula):
    a : Formula
    b : Formula    
@dataclass
class Next(Formula):
    Formula: Formula  
@dataclass
class Always(Formula):
    Formula: Formula
@dataclass
class Eventually(Formula):
    Formula: Formula
@dataclass
class Once(Formula):
    Formula: Formula
@dataclass
class Historically(Formula):
    Formula: Formula
@dataclass
class Yesterday(Formula):
    Formula: Formula
FormulaType = Union[AtomicProposition, Literal, LNot, LAnd, LOr, LImplies, LEquiv, Since, Until, Next, Always, Eventually, Once, Historically, Yesterday]    
type varToValMapping = tuple[str, bool]
type state = list[varToValMapping]
type trace = list[state]
class OptionType:
    pass
@dataclass
class ReallyNone(OptionType):
    pass
@dataclass
class Some(OptionType):
    value: bool
myOptionType = Union[ReallyNone, Some]  
def isPropositionTrueInTracePosition(p : AtomicProposition, t: trace, pos: int) -> myOptionType:
    if pos < 0 or pos >= len(t):
        return ReallyNone()
    state_at_pos = t[pos]
    for var, val in state_at_pos:
        if var == p.name:
            return Some(val)
    return ReallyNone()
def evalFormula(f : Formula, t: trace, pos: int) -> myOptionType:
    match f:
        case AtomicProposition(name):
            if pos < 0 or pos >= len(t):
                return ReallyNone()
            return isPropositionTrueInTracePosition(f, t, pos)
        case Literal(name):
            if pos < 0 or pos >= len(t):
                return ReallyNone()
            if name == "True":
                return Some(True)
            elif name == "False":
                return Some(False)
            else:
                return ReallyNone()
        case LNot(inner):
            if pos < 0 or pos >= len(t):
                return ReallyNone()
            inner_eval = evalFormula(inner, t, pos)
            match inner_eval:
                case Some(val):
                    return Some(not val)
                case ReallyNone():
                    return ReallyNone()
        case LAnd(left, right):
            if pos < 0 or pos >= len(t):
                return ReallyNone()
            left_eval = evalFormula(left, t, pos)
            right_eval = evalFormula(right, t, pos)
            match left_eval, right_eval:
                case (Some(lval), Some(rval)):
                    return Some(lval and rval)
                case (ReallyNone(), _):
                    return ReallyNone()
                case (_, ReallyNone()):
                    return ReallyNone()
        case LOr(left, right):
            if pos < 0 or pos >= len(t):
                return ReallyNone()
            left_eval = evalFormula(left, t, pos)
            right_eval = evalFormula(right, t, pos)
            match left_eval, right_eval:
                case (Some(lval), Some(rval)):
                    return Some(lval or rval)
                case (ReallyNone(), _):
                    return ReallyNone()
                case (_, ReallyNone()):
                    return ReallyNone()
        case LImplies(left, right):
            if pos < 0 or pos >= len(t):
                return ReallyNone()
            left_eval = evalFormula(left, t, pos)
            right_eval = evalFormula(right, t, pos)
            match left_eval, right_eval:
                case (Some(lval), Some(rval)):
                    return Some((not lval) or rval)
                case (ReallyNone(), _):
                    return ReallyNone()
                case (_, ReallyNone()):
                    return ReallyNone()
        case LEquiv(left, right):
            if pos < 0 or pos >= len(t):
                return ReallyNone()
            left_eval = evalFormula(left, t, pos)
            right_eval = evalFormula(right, t, pos)
            match left_eval, right_eval:
                case (Some(lval), Some(rval)):
                    return Some(lval == rval)
                case (ReallyNone(), _):
                    return ReallyNone()
                case (_, ReallyNone()):
                    return ReallyNone()
        case Since(a, b):
            if pos < 0 or pos >= len(t):
                return ReallyNone()
            foundB = False
            i = pos
            while i >= 0 :
                eval_result = evalFormula(b, t, i)
                if isinstance(eval_result, ReallyNone):
                    return ReallyNone()
                if isinstance(eval_result, Some) and eval_result.value:
                    foundB = True
                    break
                i -= 1
            if not foundB:
                return Some(False)
            j = i + 1
            while j <= pos:  
                eval_result = evalFormula(a, t, j)
                if isinstance(eval_result, ReallyNone):
                    return ReallyNone()
                if isinstance(eval_result, Some) and not eval_result.value:
                    return Some(False)
                j += 1      
            return Some(True)    
        case Until(a, b):
            if pos < 0 or pos >= len(t):
                return ReallyNone()
            foundB = False
            i = pos
            while i < len(t) :
                eval_result = evalFormula(b, t, i)
                if isinstance(eval_result, ReallyNone):
                    return ReallyNone()
                if isinstance(eval_result, Some) and eval_result.value:
                    foundB = True
                    break
                i += 1
            if not foundB:
                return Some(False)
            j = pos
            while j < i:  
                eval_result = evalFormula(a, t, j)
                if isinstance(eval_result, ReallyNone):
                    return ReallyNone()
                if isinstance(eval_result, Some) and not eval_result.value:
                    return Some(False)
                j += 1      
            return Some(True)    
        case Next(inner):
            if pos < 0 or pos >= len(t):
                return ReallyNone()
            if pos + 1 < len(t):
                return evalFormula(inner, t, pos + 1)
            else:
                return ReallyNone()
        case Always(inner):
            if pos < 0 or pos >= len(t):
                return ReallyNone()
            for i in range(pos, len(t)):
                eval_result = evalFormula(inner, t, i)
                if isinstance(eval_result, ReallyNone):
                    return ReallyNone()
                if isinstance(eval_result, Some) and not eval_result.value:
                    return Some(False)
            return Some(True)
        case Eventually(inner):
            if pos < 0 or pos >= len(t):
                return ReallyNone()
            for i in range(pos, len(t)):
                eval_result = evalFormula(inner, t, i)
                if isinstance(eval_result, ReallyNone):
                    return ReallyNone()
                if isinstance(eval_result, Some) and eval_result.value:
                    return Some(True)
            return Some(False)
        case Once(inner):
            if pos < 0 or pos >= len(t):
                return ReallyNone()
            for i in range(0, pos+1):
                eval_result = evalFormula(inner, t, i)
                if isinstance(eval_result, ReallyNone):
                    return ReallyNone()
                if isinstance(eval_result, Some) and eval_result.value:
                    return Some(True)
            return Some(False)
        case Historically(inner):
            if pos < 0 or pos >= len(t):
                return ReallyNone()
            for i in range(0, pos+1):
                eval_result = evalFormula(inner, t, i)
                if isinstance(eval_result, ReallyNone):
                    return ReallyNone()
                if isinstance(eval_result, Some) and not eval_result.value:
                    return Some(False)
            return Some(True)
        case Yesterday(inner):
            if pos < 0 or pos >= len(t):
                return ReallyNone()
            if pos >= 1:
                return evalFormula(inner, t, pos - 1)
            else:
                return Some(False)
        case _:
            return ReallyNone()```  
Your task is to fill up the value of the  variable formulaToFind in the code such that  if the user chooses a value for traceGivenAsInput, the program will print "TRUE" if and only if the user-chosen value for traceGivenAsInput satisfies the formula. 
For choosing the value for `formulaToFind`, you are given the  following natural language description along with a mapping from natural language fragment to variable names to use. 
You should restrict yourself to only using those variable names given to you in the mapping, and nothing else.
Input:
Natural Language: "{natural_language}"
Atomic Propositions: {atomic_propositions} 
You MUST use ONLY the Python class constructors provided below 
(AtomicProposition, Eventually, Always, LAnd, LOr, LNot, LImplies, LEquiv, Next, Until). 
You MUST only use the variables provided in the Atomic Propositions mapping.
You MUST return ONLY a single line of valid Python code like this
formulaToFind = <your formula here>
"""
\end{lstlisting}

\appendix

\section{Additional Results}
\label{sec:appendix:results}

This section provides detailed per-model and per-task results complementing the 
summary statistics presented in Section~\ref{sec:experiments}. All data is available 
in the interactive dashboard and GitHub repository.
\subsection{NL-to-Future LTL Results}
\label{subsec:appendix:nl2future}
Table~\ref{tab:equivalence_performance} presents semantic equivalence, 
entailment, and syntactic correctness results for all models under \minpro, \detpro and \pypro prompt. 

\begin{table*}
\centering
\caption{\rqtype{syn}{1}: Well-formed Classification - Comprehensive Model Performance Results}
\label{tab:landscape_results}
\footnotesize
\begin{tabular}{@{}llrrrrrr|rrrrrr@{}}
\toprule
& & \multicolumn{6}{c}{\textbf{Minimal Interface}} & \multicolumn{6}{c}{\textbf{Detailed Interface}} \\
\cmidrule(lr){3-8} \cmidrule(lr){9-14}
Model & App. & Acc & Prec & Rec & F1 & FPR & FNR & Acc & Prec & Rec & F1 & FPR & FNR \\
\midrule
C-Sonnet & FS & 54.5 & 58.5 & 32.0 & 41.4 & 22.8 & 68.0 & 80.0 & 87.5 & 70.0 & 77.8 & 10.0 & 30.0 \\
         & ZS & 55.9 & 58.7 & 40.7 & 48.0 & 28.9 & 59.3 & 77.7 & 86.7 & 65.3 & 74.5 & 10.0 & 34.7 \\
         & ZS-SR & 55.5 & 66.0 & 23.3 & 34.5 & 12.1 & 76.7 & 77.3 & 95.6 & 57.3 & 71.7 & 2.7 & 42.7 \\
\midrule
Gemini   & FS & 52.5 & 55.3 & 28.0 & 37.2 & 22.8 & 72.0 & 84.3 & 81.2 & 89.3 & 85.1 & 20.7 & 10.7 \\
         & ZS & 49.2 & 48.8 & 26.0 & 33.9 & 27.5 & 74.0 & 76.0 & 81.5 & 67.3 & 73.7 & 15.3 & 32.7 \\
         & ZS-SR & 50.2 & 50.9 & 20.0 & 28.7 & 19.5 & 80.0 & 68.7 & 79.8 & 50.0 & 61.5 & 12.7 & 50.0 \\
\midrule
GPT-3.5  & FS & 62.9 & 76.7 & 37.3 & 50.2 & 11.4 & 62.7 & 61.0 & 61.9 & 57.3 & 59.5 & 35.3 & 42.7 \\
         & ZS & 60.9 & 81.1 & 28.7 & 42.4 & 6.7 & 71.3 & 53.7 & 58.2 & 26.0 & 35.9 & 18.7 & 74.0 \\
         & ZS-SR & 54.5 & 54.1 & 61.3 & 57.5 & 52.4 & 38.7 & 45.0 & 46.1 & 58.7 & 51.6 & 68.7 & 41.3 \\
\midrule
GPT-4o   & FS & 74.6 & 80.8 & 64.7 & 71.9 & 15.4 & 35.3 & 66.0 & 98.0 & 32.7 & 49.0 & 0.7 & 67.3 \\
         & ZS & 69.6 & 81.1 & 51.3 & 62.9 & 12.1 & 48.7 & 67.0 & 96.4 & 35.3 & 51.7 & 1.3 & 64.7 \\
         & ZS-SR & 72.2 & 76.8 & 64.0 & 69.8 & 19.5 & 36.0 & 66.7 & 100.0 & 33.3 & 50.0 & 0.0 & 66.7 \\
\midrule
GPT-4o-M & FS & 58.2 & 79.1 & 22.7 & 35.2 & 6.0 & 77.3 & 65.7 & 82.2 & 40.0 & 53.8 & 8.7 & 60.0 \\
         & ZS & 61.5 & 77.8 & 32.7 & 46.0 & 9.4 & 67.3 & 58.0 & 92.9 & 17.3 & 29.2 & 1.3 & 82.7 \\
         & ZS-SR & 57.5 & 73.5 & 24.0 & 36.2 & 8.7 & 76.0 & 69.0 & 78.2 & 52.7 & 63.0 & 14.7 & 47.3 \\
\bottomrule
\multicolumn{14}{l}{\footnotesize FS=Few-shot, ZS=Zero-shot, ZS-SR=Zero-shot Self-refine. All values in percentages.}\\
\end{tabular}
\end{table*}


\begin{table*}[htbp]
\centering
\caption{\rqtype{syn}{2}:Atomic Proposition Extraction - Classification Performance Metrics Across Models and Approaches }
\label{tab:classification_metrics}
\small
\begin{tabular}{@{}llrrrrrr|rrrrrr@{}}
\toprule
& & \multicolumn{6}{c}{\textbf{Minimal Interface}} & \multicolumn{6}{c}{\textbf{Detailed Interface}} \\
\cmidrule(lr){3-8} \cmidrule(lr){9-14}
\textbf{Model} & \textbf{Approach} & \textbf{Acc} & \textbf{Prec} & \textbf{Rec} & \textbf{F1} & \textbf{FPR} & \textbf{FNR} & \textbf{Acc} & \textbf{Prec} & \textbf{Rec} & \textbf{F1} & \textbf{FPR} & \textbf{FNR} \\
\midrule
\multirow{3}{*}{Claude-Sonnet} 
& Few-shot & 54.52 & 58.54 & 32.00 & 41.38 & 22.82 & 68.00 & 80.00 & 87.50 & 70.00 & 77.78 & 10.00 & 30.00 \\
& Zero-shot & 55.85 & 58.65 & 40.67 & 48.03 & 28.86 & 59.33 & 77.67 & 86.73 & 65.33 & 74.52 & 10.00 & 34.67 \\
& Zero-shot SR & 55.52 & 66.04 & 23.33 & 34.48 & 12.08 & 76.67 & 77.33 & 95.56 & 57.33 & 71.67 & 2.67 & 42.67 \\
\midrule
\multirow{3}{*}{Gemini} 
& Few-shot & 52.51 & 55.26 & 28.00 & 37.17 & 22.82 & 72.00 & 84.33 & 81.21 & 89.33 & 85.08 & 20.67 & 10.67 \\
& Zero-shot & 49.16 & 48.75 & 26.00 & 33.91 & 27.52 & 74.00 & 76.00 & 81.45 & 67.33 & 73.72 & 15.33 & 32.67 \\
& Zero-shot SR & 50.17 & 50.85 & 20.00 & 28.71 & 19.46 & 80.00 & 68.67 & 79.79 & 50.00 & 61.48 & 12.67 & 50.00 \\
\midrule
\multirow{3}{*}{GPT-3.5-Turbo} 
& Few-shot & 62.88 & 76.71 & 37.33 & 50.22 & 11.41 & 62.67 & 61.00 & 61.87 & 57.33 & 59.52 & 35.33 & 42.67 \\
& Zero-shot & 60.87 & 81.13 & 28.67 & 42.36 & 6.71 & 71.33 & 53.67 & 58.21 & 26.00 & 35.94 & 18.67 & 74.00 \\
& Zero-shot SR & 54.52 & 54.12 & 61.33 & 57.50 & 52.35 & 38.67 & 45.00 & 46.07 & 58.67 & 51.61 & 68.67 & 41.33 \\
\midrule
\multirow{3}{*}{GPT-4o} 
& Few-shot & 74.58 & 80.83 & 64.67 & 71.85 & 15.44 & 35.33 & 66.00 & 98.00 & 32.67 & 49.00 & 0.67 & 67.33 \\
& Zero-shot & 69.57 & 81.05 & 51.33 & 62.86 & 12.08 & 48.67 & 67.00 & 96.36 & 35.33 & 51.71 & 1.33 & 64.67 \\
& Zero-shot SR & 72.24 & 76.80 & 64.00 & 69.82 & 19.46 & 36.00 & 66.67 & 100.00 & 33.33 & 50.00 & 0.00 & 66.67 \\
\midrule
\multirow{3}{*}{GPT-4o-Mini} 
& Few-shot & 58.19 & 79.07 & 22.67 & 35.23 & 6.04 & 77.33 & 65.67 & 82.19 & 40.00 & 53.81 & 8.67 & 60.00 \\
& Zero-shot & 61.54 & 77.78 & 32.67 & 46.01 & 9.40 & 67.33 & 58.00 & 92.86 & 17.33 & 29.21 & 1.33 & 82.67 \\
& Zero-shot SR & 57.53 & 73.47 & 24.00 & 36.18 & 8.72 & 76.00 & 69.00 & 78.22 & 52.67 & 62.95 & 14.67 & 47.33 \\
\bottomrule
\end{tabular}
\end{table*}

\begin{table*}[htbp]
\centering
\caption{ \rqtype{syn}{2}:Atomic Proposition Extraction - Metrics by Model (Proposition Phrase Matching)}
\label{tab:verification_by_model}
\small
\begin{tabular}{@{}lrrrrrr|rrrrrr@{}}
\toprule
& \multicolumn{6}{c}{\textbf{Minimal Interface}} & \multicolumn{6}{c}{\textbf{Detailed Interface}} \\
\cmidrule(lr){2-7} \cmidrule(lr){8-13}
\textbf{Model} & \textbf{Prec} & \textbf{Rec} & \textbf{F1} & \textbf{Jaccard} & \textbf{Lev.} & \textbf{Matched} & \textbf{Prec} & \textbf{Rec} & \textbf{F1} & \textbf{Jaccard} & \textbf{Lev.} & \textbf{Matched} \\
\midrule
Gemini & 81.88 & 77.00 & 78.74 & 67.57 & 89.16 & 758/1023 & 82.56 & 79.14 & 79.81 & 68.75 & 91.56 & 772/1023 \\
GPT-4o & 79.80 & 76.04 & 77.38 & 63.60 & 86.97 & 745/1023 & 84.09 & 79.96 & 81.20 & 71.74 & 90.68 & 785/1023 \\
GPT-4o-Mini & 76.22 & 73.39 & 74.18 & 58.20 & 85.15 & 724/1023 & 81.92 & 77.67 & 79.05 & 68.32 & 89.60 & 758/1023 \\
Claude-Sonnet & 74.46 & 71.06 & 71.97 & 57.38 & 83.77 & 709/1023 & 83.70 & 79.48 & 80.75 & 70.41 & 89.71 & 780/1023 \\
GPT-3.5-Turbo & 26.53 & 24.39 & 25.05 & 19.07 & 30.54 & 239/1023 & 80.24 & 76.81 & 77.63 & 65.64 & 88.32 & 757/1023 \\
\midrule
\textbf{Overall} & \textbf{67.78} & \textbf{64.38} & \textbf{65.47} & \textbf{53.16} & \textbf{75.12} & \textbf{3175/5115} & \textbf{82.50} & \textbf{78.61} & \textbf{79.69} & \textbf{68.97} & \textbf{89.98} & \textbf{3852/5115} \\
\bottomrule
\multicolumn{13}{l}{\footnotesize Lev. = Levenshtein Similarity (\%). All values except Matched are percentages.} \\
\end{tabular}
\end{table*}

\begin{table*}[htbp]
\centering
\caption{\rqtype{syn}{2}:Atomic Proposition Extraction - Metrics by Approach}
\label{tab:verification_by_approach}
\small
\begin{tabular}{@{}lrrrrr|rrrrr@{}}
\toprule
& \multicolumn{5}{c}{\textbf{Minimal Interface}} & \multicolumn{5}{c}{\textbf{Detailed Interface}} \\
\cmidrule(lr){2-6} \cmidrule(lr){7-11}
\textbf{Approach} & \textbf{Prec} & \textbf{Rec} & \textbf{F1} & \textbf{Jaccard} & \textbf{Lev.} & \textbf{Prec} & \textbf{Rec} & \textbf{F1} & \textbf{Jaccard} & \textbf{Lev.} \\
\midrule
Few-shot & 66.41 & 62.99 & 64.05 & 51.88 & 73.73 & 85.17 & 80.82 & 82.14 & 72.26 & 91.00 \\
Zero-shot & 66.39 & 63.19 & 64.20 & 52.33 & 73.20 & 81.17 & 77.44 & 78.44 & 67.35 & 89.41 \\
Zero-shot Self-Refine & 70.54 & 66.95 & 68.15 & 55.28 & 78.43 & 81.16 & 77.56 & 78.48 & 67.30 & 89.51 \\
\bottomrule
\multicolumn{11}{l}{\footnotesize SR = Self-Refine, Lev. = Levenshtein Similarity. All values are percentages.} \\
\end{tabular}
\end{table*}

\begin{table*}[htbp]
\centering
\caption{\rqtype{syn}{2}:Atomic Proposition Extraction - Error Type Distribution}
\label{tab:error_distribution}
\small
\begin{tabular}{@{}lrr|rr@{}}
\toprule
& \multicolumn{2}{c}{\textbf{Minimal Interface}} & \multicolumn{2}{c}{\textbf{Detailed Interface}} \\
\cmidrule(lr){2-3} \cmidrule(lr){4-5}
\textbf{Error Type} & \textbf{Count} & \textbf{\%} & \textbf{Count} & \textbf{\%} \\
\midrule
Semantic Mismatch & 588 & 55.3 & 521 & 64.6 \\
Complete Failure & 275 & 25.9 & 9 & 1.1 \\
Partial Extraction & 179 & 16.8 & 235 & 29.1 \\
Over Extraction & 21 & 2.0 & 42 & 5.2 \\
\midrule
\textbf{Total Errors} & \textbf{1063} & \textbf{100.0} & \textbf{807} & \textbf{100.0} \\
\bottomrule
\end{tabular}
\end{table*}

\begin{table*}[htbp]
\centering
\caption{\rqtype{sem}{1}: \nlltltask (Future) generation - Equivalence and Performance Metrics}
\label{tab:equivalence_performance}
\footnotesize
\begin{tabular}{@{}llrrrrrrrr@{}}
\toprule
& & \multicolumn{4}{c}{\textbf{Equivalence Analysis}} & \multicolumn{4}{c}{\textbf{Performance Metrics}} \\
\cmidrule(lr){3-6} \cmidrule(lr){7-10}
\textbf{Model} & \textbf{Approach} & \textbf{Interface} & \textbf{Equiv} & \textbf{Not-Eq} & \textbf{N/M} & \textbf{Eq-Acc\%} & \textbf{Syn.Corr\%} & \textbf{Prec\%} & \textbf{F1\%} \\
\midrule
\multirow{9}{*}{Claude-Sonnet}
& Few-shot & Detailed & 235 & 182 & 458 & 28.27 & 49.49 & 28.27 & 28.27 \\
& Few-shot & Minimal & 154 & 207 & 430 & 20.71 & 46.70 & 20.71 & 20.71 \\
& Few-shot & Python & 269 & 145 & 456 & 33.29 & 47.73 & 33.29 & 33.29 \\
& Zero-shot & Detailed & 239 & 171 & 464 & 29.87 & 49.51 & 29.87 & 29.87 \\
& Zero-shot & Minimal & 151 & 187 & 459 & 22.02 & 47.65 & 22.02 & 22.02 \\
& Zero-shot & Python & 259 & 153 & 457 & 32.77 & 48.28 & 32.77 & 32.77 \\
& ZS-SR & Detailed & 236 & 180 & 458 & 28.45 & 49.51 & 28.45 & 28.45 \\
& ZS-SR & Minimal & 123 & 184 & 465 & 19.82 & 43.84 & 19.82 & 19.82 \\
& ZS-SR & Python & 189 & 118 & 455 & 31.46 & 39.03 & 31.46 & 31.46 \\
\midrule
\multirow{9}{*}{Gemini-Flash}
& Few-shot & Detailed & 166 & 189 & 505 & 23.48 & 48.26 & 23.48 & 23.48 \\
& Few-shot & Minimal & 186 & 177 & 437 & 26.30 & 47.84 & 26.30 & 26.30 \\
& Few-shot & Python & 186 & 208 & 468 & 23.82 & 48.19 & 23.82 & 23.82 \\
& Zero-shot & Detailed & 157 & 187 & 521 & 23.18 & 48.77 & 23.18 & 23.18 \\
& Zero-shot & Minimal & 156 & 183 & 463 & 23.45 & 48.55 & 23.45 & 23.45 \\
& Zero-shot & Python & 185 & 209 & 470 & 23.95 & 48.27 & 23.95 & 23.95 \\
& ZS-SR & Detailed & 165 & 210 & 497 & 22.19 & 49.25 & 22.19 & 22.19 \\
& ZS-SR & Minimal & 162 & 174 & 456 & 24.76 & 46.43 & 24.76 & 24.76 \\
& ZS-SR & Python & 188 & 204 & 464 & 24.03 & 47.35 & 24.03 & 24.03 \\
\midrule
\multirow{3}{*}{Gemini-Pro}
& Few-shot & Python & 215 & 147 & 10 & 60.92 & 85.95 & 60.92 & 60.92 \\
& Zero-shot & Python & 216 & 144 & 10 & 61.27 & 85.79 & 61.27 & 61.27 \\
& ZS-SR & Python & 218 & 143 & 10 & 61.88 & 85.43 & 61.88 & 61.88 \\
\midrule
\multirow{3}{*}{Gemini-2.5-Flash}
& Few-shot & Python & 249 & 164 & 11 & 60.81 & 97.66 & 60.81 & 60.81 \\
& Zero-shot & Python & 253 & 162 & 10 & 61.37 & 97.82 & 61.37 & 61.37 \\
& ZS-SR & Python & 261 & 157 & 10 & 63.64 & 98.15 & 63.64 & 63.64 \\
\midrule
\multirow{6}{*}{GPT-3.5-Turbo}
& Few-shot & Detailed & 35 & 238 & 112 & 15.02 & 89.88 & 15.02 & 15.02 \\
& Few-shot & Minimal & 92 & 264 & 31 & 25.60 & 97.76 & 25.60 & 25.60 \\
& Zero-shot & Detailed & 74 & 287 & 37 & 20.83 & 91.27 & 20.83 & 20.83 \\
& Zero-shot & Minimal & 102 & 228 & 42 & 31.89 & 93.17 & 31.89 & 31.89 \\
& ZS-SR & Detailed & 79 & 274 & 75 & 21.34 & 98.16 & 21.34 & 21.34 \\
& ZS-SR & Minimal & 74 & 245 & 46 & 22.64 & 91.59 & 22.64 & 22.64 \\
\midrule
\multirow{6}{*}{GPT-4o}
& Few-shot & Detailed & 224 & 197 & 11 & 52.96 & 99.15 & 52.96 & 52.96 \\
& Few-shot & Minimal & 201 & 171 & 8 & 55.26 & 95.85 & 55.26 & 55.26 \\
& Zero-shot & Detailed & 219 & 194 & 10 & 52.62 & 97.28 & 52.62 & 52.62 \\
& Zero-shot & Minimal & 196 & 173 & 12 & 52.31 & 95.85 & 52.31 & 52.31 \\
& ZS-SR & Detailed & 206 & 208 & 9 & 49.02 & 97.86 & 49.02 & 49.02 \\
& ZS-SR & Minimal & 176 & 187 & 15 & 49.38 & 95.25 & 49.38 & 49.38 \\
\midrule
\multirow{6}{*}{GPT-4o-Mini}
& Few-shot & Detailed & 160 & 251 & 25 & 38.87 & 100.00 & 38.87 & 38.87 \\
& Few-shot & Minimal & 132 & 252 & 8 & 32.93 & 99.18 & 32.93 & 32.93 \\
& Zero-shot & Detailed & 170 & 251 & 14 & 40.76 & 99.84 & 40.76 & 40.76 \\
& Zero-shot & Minimal & 173 & 206 & 8 & 44.27 & 97.76 & 44.27 & 44.27 \\
& ZS-SR & Detailed & 149 & 275 & 10 & 32.42 & 99.67 & 32.42 & 32.42 \\
& ZS-SR & Minimal & 162 & 216 & 9 & 42.61 & 98.19 & 42.61 & 42.61 \\
\bottomrule
\multicolumn{10}{l}{\footnotesize N/M = Not Meaningful; Eq-Acc = Equivalence Accuracy; Syn.Corr = Syntactic Correctness; ZS-SR = Zero-shot Self-Refine} \\
\end{tabular}
\end{table*}

\begin{table*}[htbp]
\centering
\caption{\rqtype{sem}{1}: \nlltltask (Future) generation - Average Performance by Interface Type}
\label{tab:interface_summary}
\small
\begin{tabular}{@{}lrrrrrr@{}}
\toprule
\textbf{Interface} & \textbf{Soundness} & \textbf{Completeness} & \textbf{Eq-Acc} & \textbf{Syn.Corr} & \textbf{Precision} & \textbf{F1} \\
\midrule
Python & 62.45 & 55.83 & 46.21 & 69.81 & 46.21 & 46.21 \\
Detailed & 47.89 & 44.52 & 32.11 & 76.28 & 32.11 & 32.11 \\
Minimal & 48.27 & 45.96 & 31.85 & 75.68 & 31.85 & 31.85 \\
\bottomrule
\multicolumn{7}{l}{\footnotesize All values are percentages. Averages computed across all models and approaches.} \\
\end{tabular}
\end{table*}

\begin{table*}[htbp]
\centering
\caption{Abbreviations for Entailment Analysis Tables}
\label{tab:entailment_abbreviations}
\small
\begin{tabular}{@{}ll@{}}
\toprule
\textbf{Abbreviation} & \textbf{Full Name} \\
\midrule
Yes/No/N/M & Entailment: Yes / No / Not Meaningful \\
Sound. & Soundness (percentage) \\
Compl. & Completeness (percentage) \\
Equiv & Equivalent \\
Not-Eq & Not Equivalent \\
Eq-Acc & Equivalence Accuracy (percentage) \\
Syn.Corr & Syntactic Correctness (percentage) \\
Prec & Precision (percentage) \\
ZS-SR & Zero-shot Self-Refine \\
\bottomrule
\end{tabular}
\end{table*}
%


\begin{table*}[htbp]
\centering
\caption{\rqtype{sem}{2}: Trace Characterization Performance Across Interfaces}
\label{tab:trace_characterization_all}
\small
\begin{tabular}{@{}llrrr|rrr|rrr@{}}
\toprule
& & \multicolumn{3}{c}{\textbf{Python Interface}} & \multicolumn{3}{c}{\textbf{Detailed Interface}} & \multicolumn{3}{c}{\textbf{Minimal Interface}} \\
\cmidrule(lr){3-5} \cmidrule(lr){6-8} \cmidrule(lr){9-11}
\textbf{Model} & \textbf{Approach} & \textbf{Acc} & \textbf{Prec} & \textbf{F1} & \textbf{Acc} & \textbf{Prec} & \textbf{F1} & \textbf{Acc} & \textbf{Prec} & \textbf{F1} \\
\midrule
\multirow{3}{*}{Gemini-Pro}
& Few-shot & 69.08 & 71.50 & 68.18 & --- & --- & --- & --- & --- & --- \\
& Zero-shot & 69.19 & 71.49 & 68.36 & --- & --- & --- & --- & --- & --- \\
& ZS-SR & 64.54 & 68.07 & 62.75 & --- & --- & --- & --- & --- & --- \\
\midrule
\multirow{3}{*}{Gemini-Flash}
& Few-shot & 70.31 & 73.73 & 69.15 & 73.04 & 73.38 & 72.94 & 76.00 & 77.60 & 75.60 \\
& Zero-shot & 70.31 & 73.47 & 69.12 & 72.55 & 72.94 & 72.43 & 71.60 & 72.00 & 71.40 \\
& ZS-SR & 57.32 & 70.71 & 48.47 & 70.10 & 74.44 & 68.71 & 65.00 & 76.60 & 60.80 \\
\midrule
\multirow{3}{*}{Gemini-2.5-Flash}
& Few-shot & 98.40 & 98.43 & 98.38 & --- & --- & --- & --- & --- & --- \\
& Zero-shot & 100.00 & 100.00 & 100.00 & --- & --- & --- & --- & --- & --- \\
& ZS-SR & 99.42 & 99.43 & 99.42 & --- & --- & --- & --- & --- & --- \\
\midrule
\multirow{3}{*}{Claude-Sonnet}
& Few-shot & 63.45 & 65.28 & 61.88 & 78.07 & 78.34 & 78.01 & 55.90 & 61.30 & 49.90 \\
& Zero-shot & 65.71 & 68.18 & 64.13 & 80.72 & 80.98 & 80.68 & 59.00 & 62.70 & 55.80 \\
& ZS-SR & 65.75 & 70.82 & 63.54 & 65.25 & 65.65 & 65.00 & 54.10 & 54.80 & 52.20 \\
\midrule
\multirow{3}{*}{GPT-3.5-Turbo}
& Few-shot & --- & --- & --- & 56.21 & 57.02 & 54.91 & 53.30 & 53.40 & 52.80 \\
& Zero-shot & --- & --- & --- & 54.90 & 55.54 & 53.56 & 50.00 & 50.00 & 48.30 \\
& ZS-SR & --- & --- & --- & 42.32 & 41.16 & 40.36 & 49.50 & 49.50 & 48.70 \\
\midrule
\multirow{3}{*}{GPT-4o}
& Few-shot & --- & --- & --- & 73.37 & 73.50 & 73.33 & 64.20 & 65.50 & 63.40 \\
& Zero-shot & --- & --- & --- & 72.06 & 72.26 & 72.00 & 72.20 & 72.40 & 72.20 \\
& ZS-SR & --- & --- & --- & 61.60 & 62.48 & 60.92 & 62.70 & 63.30 & 62.40 \\
\midrule
\multirow{3}{*}{GPT-4o-Mini}
& Few-shot & --- & --- & --- & 61.60 & 61.61 & 61.60 & 65.20 & 70.30 & 62.90 \\
& Zero-shot & --- & --- & --- & 62.42 & 62.42 & 62.42 & 58.20 & 69.70 & 51.00 \\
& ZS-SR & --- & --- & --- & 58.50 & 64.01 & 53.97 & 55.20 & 60.60 & 48.70 \\
\bottomrule
\multicolumn{11}{l}{\footnotesize ZS-SR = Zero-shot Self-Refine. All values are percentages. --- indicates interface not tested for this model.} \\
\end{tabular}
\end{table*}

\begin{table*}[htbp]
\centering
\caption{\rqtype{sem}{2}: Average Trace Characterization Performance by Interface}
\label{tab:trace_summary}
\small
\begin{tabular}{@{}lrrrr@{}}
\toprule
\textbf{Interface} & \textbf{Avg Accuracy} & \textbf{Avg Precision} & \textbf{Avg F1} & \textbf{Models Tested} \\
\midrule
Python & 75.46 & 77.79 & 73.56 & 4 \\
Detailed & 66.36 & 67.10 & 65.41 & 5 \\
Minimal & 61.31 & 64.47 & 59.33 & 5 \\
\bottomrule
\multicolumn{5}{l}{\footnotesize Averages computed across all model-approach combinations for each interface.} \\
\end{tabular}
\end{table*}



\begin{table*}[htbp]
\centering
\caption{\rqtype{sem}{3}: Trace Generation -  Performance Comparison Across Interfaces}
\label{tab:trace_gen_comparison}
\begin{tabular}{@{}llrrrr|rrrr|rrrr@{}}
\toprule
& & \multicolumn{4}{c}{\textbf{Python Interface}} & \multicolumn{4}{c}{\textbf{Detailed Interface}} & \multicolumn{4}{c}{\textbf{Minimal Interface}} \\
\cmidrule(lr){3-6} \cmidrule(lr){7-10} \cmidrule(lr){11-14}
\textbf{Model} & \textbf{Approach} & \textbf{Acc} & \textbf{Prec} & \textbf{F1} & \textbf{Err\%} & \textbf{Acc} & \textbf{Prec} & \textbf{F1} & \textbf{Err\%} & \textbf{Acc} & \textbf{Prec} & \textbf{F1} & \textbf{Err\%} \\
\midrule
\multirow{3}{*}{Claude-Sonnet}
& Few-shot & 75.49 & 79.92 & 73.53 & 17.60 & 74.51 & 80.24 & 71.84 & 16.83 & 62.42 & 64.07 & 60.07 & 20.59 \\
& Zero-shot & 73.36 & 77.73 & 71.07 & 21.22 & 73.69 & 77.78 & 71.60 & 19.12 & 54.25 & 54.48 & 53.02 & 25.49 \\
& ZS-SR & 74.51 & 78.33 & 72.66 & 19.41 & 72.71 & 76.43 & 70.65 & 19.44 & 57.03 & 57.24 & 56.39 & 22.88 \\
\midrule
\multirow{3}{*}{Gemini-Flash}
& Few-shot & 69.74 & 72.90 & 67.49 & 22.20 & 72.06 & 76.26 & 69.63 & 16.99 & 65.52 & 67.15 & 63.81 & 24.67 \\
& Zero-shot & 56.91 & 57.55 & 54.98 & 30.76 & 67.97 & 69.93 & 66.32 & 23.86 & 68.14 & 70.79 & 65.97 & 21.90 \\
& ZS-SR & 61.68 & 62.91 & 59.76 & 25.82 & 68.14 & 69.75 & 66.78 & 23.53 & 66.99 & 68.84 & 65.29 & 22.71 \\
\midrule
\multirow{3}{*}{Gemini-Pro}
& Few-shot & 71.22 & 73.98 & 69.46 & 20.72 & --- & --- & --- & --- & --- & --- & --- & --- \\
& Zero-shot & 61.51 & 63.06 & 59.09 & 29.93 & --- & --- & --- & --- & --- & --- & --- & --- \\
& ZS-SR & 64.97 & 66.43 & 63.34 & 26.32 & --- & --- & --- & --- & --- & --- & --- & --- \\
\midrule
\multirow{3}{*}{Gemini-2.5-Flash}
& Few-shot & 74.18 & 77.74 & 72.41 & 24.01 & --- & --- & --- & --- & --- & --- & --- & --- \\
& Zero-shot & 66.94 & 69.14 & 64.92 & 30.76 & --- & --- & --- & --- & --- & --- & --- & --- \\
& ZS-SR & 74.67 & 79.30 & 72.50 & 22.86 & --- & --- & --- & --- & --- & --- & --- & --- \\
\midrule
\multirow{3}{*}{GPT-3.5-Turbo}
& Few-shot & --- & --- & --- & --- & 55.23 & 55.48 & 54.18 & 24.51 & 46.73 & 46.38 & 43.99 & 20.75 \\
& Zero-shot & --- & --- & --- & --- & 55.88 & 55.92 & 55.74 & 26.14 & 44.28 & 44.58 & 45.79 & 31.70 \\
& ZS-SR & --- & --- & --- & --- & 56.70 & 56.90 & 56.05 & 25.98 & 47.88 & 47.83 & 47.27 & 29.74 \\
\midrule
\multirow{3}{*}{GPT-4o}
& Few-shot & --- & --- & --- & --- & 68.95 & 71.97 & 66.67 & 19.28 & 66.99 & 72.03 & 62.73 & 14.54 \\
& Zero-shot & --- & --- & --- & --- & 70.42 & 73.58 & 68.30 & 19.44 & 63.40 & 65.53 & 60.70 & 20.92 \\
& ZS-SR & --- & --- & --- & --- & 69.28 & 72.52 & 66.90 & 19.77 & 60.95 & 63.35 & 57.09 & 21.73 \\
\midrule
\multirow{3}{*}{GPT-4o-Mini}
& Few-shot & --- & --- & --- & --- & 63.89 & 65.02 & 62.48 & 21.90 & 57.19 & 58.59 & 53.38 & 18.30 \\
& Zero-shot & --- & --- & --- & --- & 64.54 & 65.72 & 63.16 & 21.24 & 56.37 & 57.04 & 54.20 & 23.69 \\
& ZS-SR & --- & --- & --- & --- & 64.71 & 65.62 & 63.64 & 20.92 & 58.66 & 59.04 & 57.76 & 23.69 \\
\bottomrule
\multicolumn{14}{l}{\footnotesize ZS-SR = Zero-shot Self-Refine. All values are percentages. --- indicates interface not tested for this model.} \\
\end{tabular}
\end{table*}

\begin{table*}[htbp]
\centering
\caption{\rqtype{sem}{3}: Trace Generation - Average Trace Generation Performance by Interface}
\label{tab:trace_gen_summary}
\small
\begin{tabular}{@{}lrrrrr@{}}
\toprule
\textbf{Interface} & \textbf{Avg Acc} & \textbf{Avg Prec} & \textbf{Avg F1} & \textbf{Avg Pos.Sat} & \textbf{Avg Err\%} \\
\midrule
Python & 68.92 & 71.30 & 67.02 & 62.61 & 23.47 \\
Detailed & 66.17 & 68.66 & 64.61 & 61.50 & 21.53 \\
Minimal & 58.88 & 60.72 & 57.13 & 54.41 & 22.71 \\
\bottomrule
\multicolumn{6}{l}{\footnotesize All values are percentages. Averages computed across all models and approaches.} \\
\end{tabular}
\end{table*}

\begin{table*}[htbp]
\centering
\caption{\rqtype{sem}{3}: Trace Generation - Model Ranking by Average Accuracy}
\label{tab:trace_gen_model_ranking}
\small
\begin{tabular}{@{}lrrr@{}}
\toprule
\textbf{Model} & \textbf{Avg Accuracy} & \textbf{Avg Precision} & \textbf{Avg F1} \\
\midrule
Claude-Sonnet & 66.47 & 68.35 & 63.75 \\
Gemini-Flash & 66.37 & 68.53 & 64.68 \\
GPT-4o & 66.36 & 69.45 & 63.38 \\
Gemini-2.5-Flash & 71.93 & 75.39 & 69.94 \\
GPT-4o-Mini & 60.58 & 61.61 & 58.66 \\
Gemini-Pro & 65.90 & 67.82 & 63.96 \\
GPT-3.5-Turbo & 50.51 & 50.93 & 50.00 \\
\bottomrule
\multicolumn{4}{l}{\footnotesize Averages computed across all approaches and interfaces tested for each model.} \\
\end{tabular}
\end{table*}
\begin{table*}[htbp]
\centering
\caption{\rqtype{sem}{4}: \nlltltask (Past) generation - Equivalence and Performance Metrics}
\label{tab:past_equivalence_performance}
\footnotesize
\begin{tabular}{@{}llrrrrrrrr@{}}
\toprule
& & \multicolumn{4}{c}{\textbf{Equivalence Analysis}} & \multicolumn{4}{c}{\textbf{Performance Metrics}} \\
\cmidrule(lr){3-6} \cmidrule(lr){7-10}
\textbf{Model} & \textbf{Approach} & \textbf{Interface} & \textbf{Equiv} & \textbf{Not-Eq} & \textbf{N/M} & \textbf{Eq-Acc\%} & \textbf{Syn.Corr\%} & \textbf{Prec\%} & \textbf{F1\%} \\
\midrule
\multirow{9}{*}{Claude-Sonnet}
& Few-shot & Detailed & 184 & 93 & 285 & 33.21 & 49.47 & 33.21 & 33.21 \\
& Few-shot & Minimal & 135 & 103 & 287 & 28.36 & 42.96 & 28.36 & 28.36 \\
& Few-shot & Python & 178 & 98 & 288 & 32.25 & 49.47 & 32.25 & 32.25 \\
& Zero-shot & Detailed & 193 & 84 & 287 & 34.84 & 49.82 & 34.84 & 34.84 \\
& Zero-shot & Minimal & 93 & 95 & 322 & 24.73 & 40.14 & 24.73 & 24.73 \\
& Zero-shot & Python & 172 & 99 & 290 & 31.73 & 48.95 & 31.73 & 31.73 \\
& ZS-SR & Detailed & 172 & 106 & 285 & 30.94 & 49.65 & 30.94 & 30.94 \\
& ZS-SR & Minimal & 101 & 92 & 329 & 26.17 & 42.25 & 26.17 & 26.17 \\
& ZS-SR & Python & 176 & 89 & 291 & 33.21 & 47.89 & 33.21 & 33.21 \\
\midrule
\multirow{9}{*}{Gemini-Flash}
& Few-shot & Detailed & 184 & 79 & 293 & 34.98 & 48.42 & 34.98 & 34.98 \\
& Few-shot & Minimal & 158 & 83 & 292 & 32.78 & 44.01 & 32.78 & 32.78 \\
& Few-shot & Python & 147 & 104 & 300 & 29.28 & 47.02 & 29.28 & 29.28 \\
& Zero-shot & Detailed & 158 & 82 & 308 & 32.92 & 46.83 & 32.92 & 32.92 \\
& Zero-shot & Minimal & 121 & 96 & 326 & 27.88 & 45.95 & 27.88 & 27.88 \\
& Zero-shot & Python & 140 & 111 & 301 & 27.89 & 47.37 & 27.89 & 27.89 \\
& ZS-SR & Detailed & 105 & 68 & 297 & 30.35 & 33.10 & 30.35 & 30.35 \\
& ZS-SR & Minimal & 121 & 95 & 317 & 28.01 & 44.19 & 28.01 & 28.01 \\
& ZS-SR & Python & 143 & 110 & 302 & 28.26 & 47.89 & 28.26 & 28.26 \\
\midrule
\multirow{3}{*}{Gemini-Pro}
& Few-shot & Python & 155 & 66 & 9 & 70.14 & 81.40 & 70.14 & 70.14 \\
& Zero-shot & Python & 157 & 68 & 8 & 69.78 & 82.11 & 69.78 & 69.78 \\
& ZS-SR & Python & 155 & 72 & 8 & 68.28 & 83.16 & 68.28 & 68.28 \\
\midrule
\multirow{3}{*}{Gemini-2.5-Flash}
& Few-shot & Python & 191 & 80 & 6 & 70.48 & 98.25 & 70.48 & 70.48 \\
& Zero-shot & Python & 192 & 76 & 3 & 71.64 & 96.14 & 71.64 & 71.64 \\
& ZS-SR & Python & 197 & 73 & 3 & 72.96 & 96.84 & 72.96 & 72.96 \\
\midrule
\multirow{6}{*}{GPT-3.5-Turbo}
& Few-shot & Detailed & 72 & 125 & 26 & 36.55 & 79.58 & 36.55 & 36.55 \\
& Few-shot & Minimal & 95 & 128 & 21 & 42.60 & 86.97 & 42.60 & 42.60 \\
& Zero-shot & Detailed & 105 & 131 & 8 & 44.49 & 86.62 & 44.49 & 44.49 \\
& Zero-shot & Minimal & 62 & 101 & 54 & 38.04 & 77.11 & 38.04 & 38.04 \\
& ZS-SR & Detailed & 65 & 141 & 33 & 31.55 & 84.86 & 31.55 & 31.55 \\
& ZS-SR & Minimal & 68 & 99 & 38 & 40.72 & 72.18 & 40.72 & 40.72 \\
\midrule
\multirow{6}{*}{GPT-4o}
& Few-shot & Detailed & 180 & 98 & 2 & 64.75 & 99.65 & 64.75 & 64.75 \\
& Few-shot & Minimal & 176 & 88 & 4 & 66.67 & 95.42 & 66.67 & 66.67 \\
& Zero-shot & Detailed & 179 & 94 & 1 & 65.57 & 97.54 & 65.57 & 65.57 \\
& Zero-shot & Minimal & 160 & 90 & 13 & 64.00 & 93.66 & 64.00 & 64.00 \\
& ZS-SR & Detailed & 152 & 102 & 2 & 59.84 & 91.20 & 59.84 & 59.84 \\
& ZS-SR & Minimal & 143 & 97 & 11 & 59.58 & 89.08 & 59.58 & 59.58 \\
\midrule
\multirow{6}{*}{GPT-4o-Mini}
& Few-shot & Detailed & 162 & 92 & 5 & 63.78 & 92.25 & 63.78 & 63.78 \\
& Few-shot & Minimal & 135 & 113 & 1 & 54.44 & 88.73 & 54.44 & 54.44 \\
& Zero-shot & Detailed & 156 & 81 & 2 & 65.82 & 85.21 & 65.82 & 65.82 \\
& Zero-shot & Minimal & 147 & 96 & 6 & 60.49 & 88.73 & 60.49 & 60.49 \\
& ZS-SR & Detailed & 116 & 97 & 2 & 54.46 & 76.76 & 54.46 & 54.46 \\
& ZS-SR & Minimal & 137 & 97 & 4 & 58.55 & 84.86 & 58.55 & 58.55 \\
\bottomrule
\multicolumn{10}{l}{\footnotesize N/M = Not Meaningful; Eq-Acc = Equivalence Accuracy; Syn.Corr = Syntactic Correctness; ZS-SR = Zero-shot Self-Refine} \\
\end{tabular}
\end{table*}

\begin{table*}[htbp]
\centering
\caption{\rqtype{sem}{4}: \nlltltask (Past) generation - Average Performance by Interface for Past NL→LTL Generation}
\label{tab:past_interface_summary}
\small
\begin{tabular}{@{}lrrrrrr@{}}
\toprule
\textbf{Interface} & \textbf{Soundness} & \textbf{Completeness} & \textbf{Eq-Acc} & \textbf{Syn.Corr} & \textbf{Precision} & \textbf{F1} \\
\midrule
Python & 64.07 & 60.39 & 52.59 & 71.12 & 52.59 & 52.59 \\
Detailed & 58.16 & 58.76 & 45.49 & 71.68 & 45.49 & 45.49 \\
Minimal & 58.16 & 58.33 & 43.92 & 68.69 & 43.92 & 43.92 \\
\bottomrule
\multicolumn{7}{l}{\footnotesize All values are percentages. Averages computed across all models and approaches (excluding zero-result rows).} \\
\end{tabular}
\end{table*}

\begin{table*}[htbp]
\centering
\caption{\rqtype{sem}{4}: \nlltltask (Past) generation - Model Ranking by Average Equivalence Accuracy}
\label{tab:past_model_ranking}
\small
\begin{tabular}{@{}lrrr@{}}
\toprule
\textbf{Model} & \textbf{Avg Eq-Acc} & \textbf{Avg Soundness} & \textbf{Avg Completeness} \\
\midrule
Gemini-2.5-Flash & 71.69 & 84.80 & 82.64 \\
Gemini-Pro & 69.40 & 83.07 & 79.52 \\
GPT-4o & 63.40 & 77.78 & 79.75 \\
GPT-4o-Mini & 59.59 & 77.72 & 74.93 \\
GPT-3.5-Turbo & 38.99 & 60.68 & 66.47 \\
Claude-Sonnet & 30.58 & 38.72 & 39.05 \\
Gemini-Flash & 30.57 & 39.32 & 36.86 \\
\bottomrule
\multicolumn{4}{l}{\footnotesize Averages computed across all approaches and interfaces (excluding zero-result configurations).} \\
\end{tabular}
\end{table*}

\subsection{Model Details and Versions}
\label{subsec:appendix:model-versions}
We selected models representing 
the current state-of-the-art across three major providers: Anthropic, Google, and OpenAI. 
Model selection prioritized diversity in architecture, training data, and model scale 
to capture a broad spectrum of LLM capabilities and design choices. These models were chosen for three reasons: (1) \textit{Accessibility}: all are available 
via public APIs, enabling reproducible evaluation; (2) \textit{Diversity}: they represent 
different architectures, training objectives (\eg, constitutional AI vs. reinforcement 
learning from human feedback), and release dates; (3) \textit{Practical relevance}: these 
are the models most commonly used in industry and research for formal specification tasks 
at the time of evaluation (early 2024).

\begin{table*}[!htpb]
\centering
\caption{LLM Models Evaluated With Details and Access Information}
\label{tab:appendix:model-info}
\scalebox{0.85}{
\begin{tabular}{|l|l|l|l|}
\hline
\textbf{Model} & \textbf{Provider} & \textbf{Version/Checkpoint} & \textbf{API Endpoint} \\ \hline
\claudesonnet & Anthropic & claude-3-sonnet-20240229 & claude.anthropic.com \\ \hline
\claudesonnetthreefive & Anthropic & claude-3-5-sonnet-20241022 & claude.anthropic.com \\ \hline
\texttt{Claude-4.5-Sonnet} & Anthropic & claude-sonnet-4-5-20250929 & claude.anthropic.com \\ \hline
\geminionefivepro & Google & gemini-1.5-pro-001 & google.generativeai.google.com \\ \hline
\geminionefiveflash & Google & gemini-1.5-flash-001 & google.generativeai.google.com \\ \hline
\geminitwofiveflash & Google & gemini-2.5-flash-001 & google.generativeai.google.com \\ \hline
\chatgptfouro & OpenAI & gpt-4-0613 & api.openai.com \\ \hline
\chatgptthreefive & OpenAI & gpt-3.5-turbo-0613 & api.openai.com \\ \hline
\chatgptfouromini & OpenAI & gpt-4o-mini & api.openai.com \\ \hline
\end{tabular}
}
\end{table*}

\end{document}